\documentclass{article}

\usepackage{arxiv}

\usepackage[utf8]{inputenc}
\usepackage[T1]{fontenc}
\usepackage[hypertexnames=false, hidelinks]{hyperref}
\usepackage{url}
\usepackage{booktabs}
\usepackage{amsfonts} 
\usepackage{nicefrac}
\usepackage{microtype}

\usepackage{amsmath}
\usepackage[labelfont=bf]{caption}
\usepackage{cite}
\usepackage[nameinlink]{cleveref}
\usepackage[]{listings}
\usepackage{empheq}
\usepackage{float}
\usepackage{graphicx}
\usepackage{siunitx}
\usepackage{xcolor}

\crefname{lstlisting}{listing}{listings}
\Crefname{lstlisting}{Listing}{Listings}

\DeclareSIUnit\pixel{px}
\DeclareSIUnit\voxel{vx}
\DeclareSIUnit\year{yr}

\DeclareMathOperator{\Tr}{Tr}

\title{Deep Learning for Reaction-Diffusion Glioma Growth Modelling: Towards a Fully Personalised Model?}

\author{
    Corentin~Martens\\
    Department of Nuclear Medicine -- H\^opital Erasme\\
    Université libre de Bruxelles\\
    Brussels, Belgium\\
    \texttt{corentin.martens@ulb.ac.be}\\
    \And
    Antonin~Rovai\\
    Department of Nuclear Medicine -- H\^opital Erasme\\
    Université libre de Bruxelles\\
    Brussels, Belgium\\
    \And
    Daniele~Bonatto\\
    Laboratory of Image Synthesis and Analysis\\
    Universit\'e libre de Bruxelles\\
    Brussels, Belgium\\
    \And
    Thierry~Metens\\
    Department of Radiology -- H\^opital Erasme\\
    Université libre de Bruxelles\\
    Brussels, Belgium\\
    \And
    Olivier~Debeir\\
    Laboratory of Image Synthesis and Analysis\\
    Universit\'e libre de Bruxelles\\
    Brussels, Belgium\\
    \And
    Christine~Decaestecker\\
    Laboratory of Image Synthesis and Analysis\\
    Universit\'e libre de Bruxelles\\
    Brussels, Belgium\\
    \And
    Serge~Goldman\\
    Department of Nuclear Medicine -- H\^opital Erasme\\
    Université libre de Bruxelles\\
    Brussels, Belgium\\
    \And
    Gaetan~Van Simaeys\\
    Department of Nuclear Medicine -- H\^opital Erasme\\
    Université libre de Bruxelles\\
    Brussels, Belgium\\
}

\begin{document}

\setlength{\abovedisplayskip}{20pt}
\setlength{\abovedisplayshortskip}{20pt}
\setlength{\belowdisplayskip}{20pt}
\setlength{\belowdisplayshortskip}{20pt}

\maketitle

\begin{abstract}
Reaction-diffusion models have been proposed for decades to capture the growth of gliomas, the most common primary brain tumours. However, severe limitations regarding the estimation of the initial conditions and parameter values of such models have restrained their clinical use as a personalised tool. In this work, we investigate the ability of deep convolutional neural networks (DCNNs) to address the pitfalls commonly encountered in the field. Based on 1,200 synthetic tumours grown over real brain geometries derived from magnetic resonance (MR) data of 6 healthy subjects, we demonstrate the ability of DCNNs to reconstruct a whole tumour cell density distribution from only two imaging contours at a single time point. With an additional imaging contour extracted at a prior time point, we also demonstrate the ability of DCNNs to accurately estimate the individual diffusivity and proliferation parameters of the model. From this knowledge, the spatio-temporal evolution of the tumour cell density distribution at later time points can ultimately be precisely captured using the model. We finally show the applicability of our approach to MR data of a real glioblastoma patient. This approach may open the perspective of a clinical application of reaction-diffusion growth models for tumour prognosis and treatment planning.
\end{abstract}


\section{Introduction}
\label{sec:1}
Gliomas are the most common primary brain tumours and remain associated with a poor prognosis. Among them, diffuse glioma are known to be highly infiltrative tumours. This, combined with the limited sensibility of magnetic resonance imaging (MRI)---the modality of choice for glioma imaging---, makes the delineation of such tumours tedious and often leads to sub-optimal treatment plannings.

Reaction-diffusion tumour growth models have been studied for decades to circumvent the limitations of current medical imaging techniques and improve treatment planning in gliomas \cite{tracqui_1995, woodward_1996, clatz_2005, hogea_2007, swanson_2008, konukoglu_2010a, konukoglu_2010b, unkelbach_2014}. These models rely on partial differential equations (PDEs) to capture the spatio-temporal evolution of a tumour cell density function over the brain domain, driven by tumour cell migration and proliferation. The most commonly used form involves a logistic reaction term and is referred to as the Fisher equation \cite{fisher_1937}:
\begin{equation}
    \frac{\partial c(\boldsymbol{r}, t)}{\partial t} = d \, \boldsymbol{\nabla}^2 c(\boldsymbol{r}, t) + \rho \, c(\boldsymbol{r}, t) \left( 1-c(\boldsymbol{r}, t) \right)
    \label{eq:1}
\end{equation}
where $c(\boldsymbol{r}, t)$ is the tumour cell density at location $\boldsymbol{r}$ and time $t$ normalised by the maximum carrying capacity $c_{\rm max}$ of brain tissues (i.e. $c(\boldsymbol{r}, t) \in [0, 1], \, \forall \boldsymbol{r},t$), $d$ is the tumour cell diffusion coefficient, and $\rho$ is the tumour cell proliferation rate. A property of the well-studied \Cref{eq:1} is that, under certain conditions and for constant coefficients, it admits a travelling wave solution on the infinite cylinder with propagation speed $v=2\sqrt{d \, \rho}$, whose profile decays exponentially with decay constant $\lambda = \sqrt{d/\rho}$ as the distance $x$ to the origin tends to infinity and for large but finite times $t$ \cite{cencini_2003, konukoglu_2010b}.

Since their first introduction by Murray and colleagues in the early 1990's \cite{tracqui_1995}, reaction-diffusion growth models have been continuously improved to successively integrate (i) a variable tumour cell diffusion rate in white versus grey matter \cite{swanson_2000} and (ii) an anisotropic diffusion tensor field accounting for the preferred migration of tumour cells along white matter tracts, whose orientation can be assessed by diffusion tensor imaging (DTI) \cite{jbabdi_2005}. These improvements led to the formulation that is used throughout this work, presented \Cref{subsec:2_1}. Tumour-induced mass effect \cite{clatz_2005, hogea_2007}, necrosis, and hypoxia \cite{swanson_2011, gu_2012}, as well as the effects of surgery \cite{woodward_1996}, chemotherapy \cite{tracqui_1995, swanson_2002, hormuth_2021}, and radiotherapy \cite{rockne_2010, hormuth_2021} have also been introduced into reaction-diffusion glioma growth models, but are not considered in this work. For a more detailed overview of reaction-diffusion glioma growth modelling, the reader is referred to \cite{tracqui_1995, woodward_1996, clatz_2005, hogea_2007, swanson_2008, konukoglu_2010a, konukoglu_2010b, unkelbach_2014}.

Although reaction-diffusion models have shown promising results for patient follow-up and improved radiotherapy planning \cite{unkelbach_2014}, their clinical application is still prone to severe limitations. Indeed, the estimation of the parameters and initial condition---i.e. the tumour cell density distribution at diagnosis time---of such models, as well as their validation in vivo, require to extract information on the tumour cell density distribution from medical imaging data. To address this issue, Swanson and colleagues have proposed in \cite{swanson_2008} to model the imaging function of MRI sequences $I_{\rm seq}(\boldsymbol{r}, t)$---indicating whether a tumour-induced abnormality is visible at location $\boldsymbol{r}$ and time $t$ on the image---as a simple tumour cell density threshold function:
\begin{equation}
    I_{\rm seq}(\boldsymbol{r}, t) =
    \begin{cases}
        1 \quad & \text{if } c(\boldsymbol{r}, t) \geq c_{\rm seq} \\
        0 \quad & \text{otherwise}
    \end{cases} 
    \label{eq:2} 
\end{equation}
where $c_{\rm seq}$ is the tumour cell density detectability threshold of the sequence. The abnormalities considered in \cite{swanson_2008} were the hyper-intense enhancing tumour core visible on T1-weighted sequences with injection of gadolinium-based contrast agent (T1Gd) and peritumour vasogenic oedema visible on T2-weighted sequences with or without fluid-attenuated inversion-recovery (T2/T2 FLAIR). Based on these assumptions, the authors suggested that the outlines of these abnormalities would correspond to iso-contours of the tumour cell density function:
    \begin{align}
    c(\boldsymbol{r}, t) = c_{\rm seq}, \quad \forall \boldsymbol{r} \in \partial \Omega_{\rm abn}
    \label{eq:3}
\end{align}
where $\partial \Omega_{\rm abn}$ is the boundary of the visible abnormality. 

Building upon this work, Konukoglu and colleagues proposed in \cite{konukoglu_2010b} a fast marching approach to construct an approximate solution of \Cref{eq:1} at imaging time which satisfies \Cref{eq:3}. This approach has the interesting property of not attempting to dynamically solve the model but seeks to extrapolate the tumour invasion beyond its MR-visible margins within the reaction-diffusion framework, with applications for radiotherapy planning. It has nevertheless two major limitations: First, it requires to be able to extract a tumour cell density iso-contour from the image, from which the whole tumour cell distribution is built. However, we showed in a previous work based on histological data that the outlines of the oedema visible on T2 FLAIR MR images do not coincide with a cell density iso-contour \cite{martens_2021}. The proposed explanation is that, due to spatial discontinuities of the tumour cell density function at interfaces between white and grey matter as well as along the brain domain boundary, \Cref{eq:2} does not necessarily imply \Cref{eq:3}. The second limitation of this approach is that the method still requires to specify the diffusivity and proliferation rate of the tumour, which are unknown at imaging time and need to be adjusted to each tumour.

The assessment of the model parameter values from medical imaging data has also been addressed previously. In \cite{konukoglu_2007}, the definition of the asymptotic speed of the tumour front $v=2\sqrt{d \, \rho}$ is used to estimate the tumour cell diffusivity $d_{\rm white}$ and $d_{\rm grey}$ in white and grey matter using a fast marching approach. However, the method does not allow to separate the individual contributions of $d$ and $\rho$ to $v$, hence $\rho$ is supposed constant for all tumours. Furthermore, this estimation is only valid for large times for which the travelling-wave approximation holds. The approach was then further extended in \cite{konukoglu_2010a} to take into account the transient speed evolution and the curvature of the tumour front, but still considers a constant $\rho$ value for all tumours. Besides, these fast marching formulations make the assumption that the outlines of the peritumour vasogenic oedema visible on T2 MR images correspond to an iso-contour of the travelling wave arrival time function. However, this hypothesis might not be verified due to discontinuities appearing at the brain boundary voxels, which could have been reached long before the imaging time. In \cite{le_2015}, a Bayesian approach is used to estimate both the diffusivity and proliferation rate parameters of the model from two imaging contours obtained by \Cref{eq:2} at two different times. The method was found to accurately estimate the infiltration length $\lambda = \sqrt{d/\rho}$ of the tumour but less accurately the tumour front propagation velocity $v=2\sqrt{d \, \rho}$, based on synthetic and real glioblastoma (GBM) MRI data. In \cite{hormuth_2021}, parameter values of a 2-species reaction-diffusion model incorporating tumour-induced mass effect and response to chemoradiation are estimated based on tumour cell density distributions derived from longitudinal T1Gd, T2 FLAIR, and diffusion-weighted (DW) MR data, with promising results. However, the cell density distributions used to initialise the model and fit the parameters were built piecewise from the enhancing/non-enhancing tumour regions delineated on T1Gd/T2 FLAIR images as well as average diffusion coefficient (ADC) maps derived from DW-MR data and are therefore not guaranteed to be solution of \Cref{eq:1} nor to reflect the actual tumour cell distribution. 

The tumour source localisation is another widely addressed problem in reaction-diffusion glioma growth modelling. In \cite{jaroudi_2016}, an inverse problem approach is used to estimate the tumour source location from a given tumour cell density distribution with promising results. However, to be applicable in clinical practice, the method still requires the ability to derive a whole tumour cell density distribution from medical imaging data.

Finally, several works have attempted to jointly estimate the tumour source location and model parameters from patient imaging data. In \cite{hogea_2008}, a PDE-constrained optimisation approach is used to assess the source location and parameter values of a reaction-diffusion glioma growth model including an additional advection term. The tumour growth model is coupled to a linear elastic model for the tumour-induced mass effect. Two optimisation criteria are used in the study: (i) the squared difference between the true and estimated cell density fields at given imaging times and (ii) the squared distance between the true and estimate position of manually defined landmarks on staggered scans, that are displaced as the surrounding brain tissues are deformed under mass effect. However, the first criterion requires the knowledge of the true tumour cell density field, which again cannot be derived directly from imaging data. Promising results were obtained for the landmark-based criterion on a real glioma case but strong assumptions are made on the initial cell density field---supposedly Gaussian---and no ground truth was available to assess the model parameter estimation. In \cite{rekik_2013}, the fast marching approach of Konukoglu and colleagues \cite{konukoglu_2007, konukoglu_2010a} is used to assess the diffusivity ratio $d_{\rm white}/d_{\rm grey}$ along with the tumour source location, but a fixed proliferation rate $\rho$ was again considered. More recently, a Bayesian framework has been proposed to simultaneously estimate the tumour source, emergence time, diffusivity, and proliferation rate from a combination of T1Gd, T2 FLAIR, and [\textsuperscript{18}F]Fluoroethyl-L-Tyrosine ([\textsuperscript{18}F]FET) positron emission tomography (PET) images in \cite{lipkova_2019}. However, the study reported that these last three parameters cannot be individually assessed from a single imaging time point.

However, none of the aforementioned works have jointly addressed the estimation of the initial condition and individual diffusivity and proliferation rate parameters of the model. Besides, most of these works considered a spatially constant diffusivity coefficient in white matter and an identical proliferation rate for all tumours. The introduction of an arbitrary diffusion tensor field $\boldsymbol{D}(\boldsymbol{r})$ and tumour-specific proliferation rate $\rho$ would make the addressed problems even more challenging.

Over the last five years, the advent of deep learning techniques---and in particular deep convolutional neural networks (DCNNs)---has opened tremendous opportunities in the field of medical imaging, achieving state-of-the-art performance in many image classification and segmentation challenges \cite{altaf_2019}. One interesting property of deep neural networks is their ability to approximate any function under certain conditions \cite{hornik_1989}. This property makes the technique attractive to address the aforementioned limitations. DCNNs may indeed be used to approximate solutions of PDEs such as \Cref{eq:1} over complex domains and for spatially variable coefficients, as well as to assess their parameter values, from partial observations provided in the form of thresholding contours.

In this work, we investigate the ability of DCNNs to address common pitfalls encountered in the clinical application of reaction-diffusion glioma growth models. In particular, we focus on the following two tasks:
\begin{enumerate}
    \item Reconstructing a whole brain tumour cell density distribution compatible with the reaction-diffusion model from a pair of contours obtained through a threshold-like imaging process as in \Cref{eq:2} for two different detectability threshold values at a given imaging time. These contours may for example correspond to the outlines of the enhancing core and peritumour vasogenic oedema on T1Gd and T2/T2 FLAIR MR images, respectively.
    \item Estimating the values of the diffusion and proliferation parameters of the model from three imaging contours: (i) two contours obtained for a first detectability threshold value (e.g. the vasogenic oedema outlines) at two different imaging times and (ii) a third contour obtained for a second detectability threshold value (e.g. the enhancing core outlines) at the second imaging time.
\end{enumerate}
We demonstrate the ability of DCNNs to perform these tasks accurately based on 1,200 synthetic tumours grown over the brain geometries derived from the MR data of 6 healthy subjects. We also show the applicability of our approach on MR data of a real glioblastoma patient.

\section{Methods}
The following sections successively describe the reaction-diffusion model considered, the MR image acquisition and processing steps, the tumour dataset synthesis, the architecture and training procedure of the DCNNs, and the validation methodology adopted for our approach.

\subsection{The Reaction-Diffusion Model}
\label{subsec:2_1}
The reaction-diffusion tumour growth model that is used throughout this work is described by the following set of equations \cite{clatz_2005,rekik_2013,stretton_2013}:
\begin{empheq}[left=\empheqlbrace\;]{align}
    &\frac{\partial c(\boldsymbol{r}, t)}{\partial t} = \boldsymbol{\nabla} \cdot \left( \boldsymbol{D}(\boldsymbol{r}) \, \boldsymbol{\nabla} c(\boldsymbol{r}, t) \right) + \rho \, c(\boldsymbol{r}, t) \left( 1-c(\boldsymbol{r}, t) \right) & \forall \boldsymbol{r} \in \Omega, \; \forall t > 0
    \label{eq:4} \\ 
    &c(\boldsymbol{r}, 0) = c_0(\boldsymbol{r}) & \forall \boldsymbol{r} \in \Omega
    \label{eq:5} \\ 
    &\boldsymbol{D}(\boldsymbol{r}) \, \boldsymbol{\nabla} c(\boldsymbol{r}, t) \cdot \boldsymbol{n}_{\partial \Omega}(\boldsymbol{r}) = 0 & \forall \boldsymbol{r} \in \partial \Omega
    \label{eq:6}
\end{empheq}
where $c(\boldsymbol{r}, t)$ is the tumour cell density at location $\boldsymbol{r}$ and time $t$ normalised by the maximum carrying capacity $c_{\rm max}$ of brain tissues (i.e. $c(\boldsymbol{r}, t) \in [0, 1], \, \forall \boldsymbol{r},t$), $\boldsymbol{D}(\boldsymbol{r})$ is the symmetric tumour cell diffusion tensor at location $\boldsymbol{r}$, $\rho$ is the tumour cell proliferation rate, $c_0(\boldsymbol{r})$ is the initial tumour cell density at location $\boldsymbol{r}$, and $\boldsymbol{n}_{\partial_{\Omega}}(\boldsymbol{r})$ is a unit normal vector pointing outwards the boundary $\partial_{\Omega}$ of the brain domain $\Omega$ at location $\boldsymbol{r} \in \partial_{\Omega}$. \Cref{eq:5} specifies the initial condition of the problem whereas \Cref{eq:6} provides no-flux Neumann boundary conditions reflecting the inability of tumour cells to diffuse across $\partial_{\Omega}$.

\subsection{MR Data Acquisition}
\label{subsec:2_2}
For the needs of this work, 6 healthy volunteers were enrolled for an MRI acquisition comprising a T1 BRAVO (echo time: $\sim \!\! \SI{3}{\milli \second}$, repetition time: $\sim \!\! \SI{8.3}{\milli \second}$, inversion time: $\SI{450}{\milli \second}$, flip angle: \SI{12}{\degree}, pixel bandwidth: \SI{244}{\hertz \per \pixel}, slice thickness/spacing: \SI{1/1}{\milli \meter}, matrix: $\SI{240}{\pixel} \times \SI{240}{\pixel}$), a T2 FLAIR (echo time: $\sim \!\! \SI{119}{\milli \second}$, repetition time: \SI{7.2}{\second}, inversion time: $\sim \!\! \SI{2040}{\milli \second}$, flip angle: \SI{90}{\degree}, bandwidth: \SI{122}{\hertz \per \pixel}, phase/slice acceleration factor: 2/2, slice thickness/spacing: \SI{1.4/0.7}{\milli \meter}, matrix: $\SI{256}{\pixel} \times \SI{256}{\pixel}$), and an EPI-DTI (echo time: \SI{77.1}{\milli \second}, repetition time: \SI{7}{\second}, inversion time: \SI{108}{\milli \second}, flip angle: \SI{90}{\degree}, bandwidth: \SI{1953.12}{\hertz \per \pixel}, phase/slice acceleration factor: 2/1, multiband factor: 3, slice thickness/spacing: \SI{2/2}{\milli \meter}, matrix: $\SI{120}{\pixel} \times \SI{120}{\pixel}$, directions: 32, $b$-value: \SI{1000}{\second \per \square \milli \meter}) sequence. To correct for susceptibility-induced distortions (see \Cref{subsubsec:2_3_1}), a second DTI acquisition with reversed phase-encode polarity and only 6 directions was additionally performed. All acquisitions were performed on a \SI{3}{\tesla} Signa PET/MR scanner (GE Healthcare, USA) with a Nova 32-channel head coil (Nova Medical, USA).

To validate our approach (see \Cref{subsec:2_6}), similar T1 BRAVO, T2 FLAIR, and DTI images as well as an additional T1Gd (echo time: $\SI{3.2}{\milli \second}$, repetition time: $\SI{8}{\milli \second}$, flip angle: \SI{8}{\degree}, pixel bandwidth: \SI{255}{\hertz \per \pixel}, slice thickness/spacing: \SI{1/1}{\milli \meter}, matrix: $\SI{232}{\pixel} \times \SI{231}{\pixel}$) image acquired on a \SI{3}{\tesla} Achieva scanner (Philips Medical Systems, The Netherlands) of a 69-year-old male patient with IDH-wildtype GBM were retrospectively analysed.

The study was conducted according to the guidelines of the Declaration of Helsinki and approved by the Hospital-Faculty Ethics Committee of H\^opital Erasme (accreditation 021/406, protocol P2018/311, 3 May 2018). Informed consent was obtained from all subjects involved in the study.

\subsection{Processing}
\label{subsec:2_3}
\subsubsection{DTI Data Processing}
\label{subsubsec:2_3_1}
The acquired DTI data were first corrected for susceptibility-induced distortion, eddy currents, and patient motion using the \lstinline{topup} and \lstinline{eddy} tools available as part of FSL \cite{jenkinson_2012}. A water diffusion tensor field $\boldsymbol{D}_{\rm water}$ was then reconstructed from the corrected DTI data by least-squares fitting using FSL's \lstinline{dtifit} tool. The whole FSL script used for DTI data processing is available in \Cref{app:a}.

\subsubsection{Resampling and Registration}
The acquired T1 BRAVO, T1Gd (GBM patient only), and T2 FLAIR images as well as the corrected DTI data and the derived water diffusion tensor field were resampled to an isotropic voxel size of $\SI{1}{\milli \metre} \times \SI{1}{\milli \metre} \times \SI{1}{\milli \metre}$ by linear interpolation. To correct for patient motion throughout the acquisition, the T1 BRAVO and T2 FLAIR images were rigidly registered to the $b=0$ DTI image used as reference by maximisation of their mutual information \cite{maes_1997}. All resampling and registration steps were performed using an in-house software in C++ based on ITK \cite{yoo_2002} and VTK \cite{schroeder_2010}.

\subsubsection{Skull Stripping}
The brain volume was then segmented on the registered T2 FLAIR image using the Otsu thresholding \cite{otsu_1979} followed by a morphological erosion with structuring element radius 2 voxels (\si{\voxel}), a largest component extraction, a morphological dilation of radius \SI{2}{\voxel}, a morphological closing of radius \SI{8}{\voxel}, and a morphological dilation of radius \SI{1}{\voxel}.

\subsubsection{Brain Tissue Segmentation}
\label{subsubsec:2_3_4}
\begin{figure}[ht!]
\centering
\includegraphics[width=0.75\textwidth]{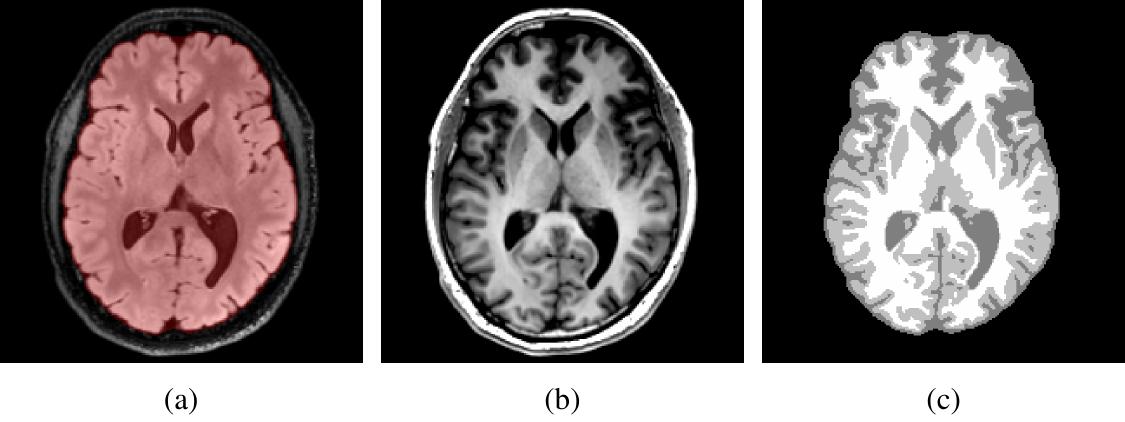}
\caption{Example of processed MR data. \textbf{(a)} Axial slice of the T2 FLAIR image with superimposed segmented brain mask (red). \textbf{(b)} Corresponding slice of the T1 BRAVO image. \textbf{(c)} Segmented brain map obtained with the MICO algorithm followed by manual corrections.}
\label{fig:1}
\end{figure}

The extracted brain domain was further segmented on the registered T1 BRAVO image into white matter, grey matter, and cerebrospinal fluid using an in-house implementation of the MICO intensity-based clustering algorithm comprising a bias field correction step \cite{li_2014}. Manual corrections were further applied to the mis-segmented basal nuclei and falx cerebri. This last step is crucial to prevent the migration of tumour cells between brain hemispheres via other routes than the corpus callosum, as highlighted previously \cite{unkelbach_2014}. The segmentation results were finally merged into a single brain map. An example of T2 FLAIR and T1 BRAVO images with the corresponding brain mask and segmented brain map is depicted in \Cref{fig:1}.

\subsubsection{Tumour Segmentation}
\label{subsubsec:2_3_5}
For validation purpose (see \Cref{subsec:2_6}), the enhancing core and peritumour vasogenic oedema were also semi-automatically segmented on the T1Gd and T2 FLAIR images of the GBM patient using combinations of thresholding, connected component extraction, and morphological operations.

\subsubsection{Tumour Diffusion Tensor}
\label{subsubsec:2_3_6}
A tumour diffusion tensor field $\boldsymbol{D}(\boldsymbol{r})$ was built piecewise from the DTI-derived water diffusion tensor $\boldsymbol{D}_{\rm water}(\boldsymbol{r})$ and the segmented brain domain $\Omega$ as follows:
\begin{equation}
    \boldsymbol{D}(\boldsymbol{r}) = 
    \begin{cases}
        \boldsymbol{D}_{\rm white}(\boldsymbol{r}) \quad & \text{if } \boldsymbol{r} \in \Omega_{\rm white} \\
        \boldsymbol{D}_{\rm grey} \quad & \text{if } \boldsymbol{r} \in \Omega_{\rm grey} \\
        \boldsymbol{0} \quad & \text{otherwise}
    \end{cases}
    \label{eq:7}
\end{equation}
where $\boldsymbol{D}_{\rm white}(\boldsymbol{r})$ and $\boldsymbol{D}_{\rm grey}$ are respectively the tumour cell diffusion tensor fields within the white and grey matter domains $\Omega_{\rm white}$ and $\Omega_{\rm grey}$, and $\boldsymbol{0}$ is the null tensor, with:
\begin{equation}
    \boldsymbol{D}_{\rm grey} = 
    \begin{bmatrix}
    d_{\rm grey} & 0 & 0 \\
    0 & d_{\rm grey} & 0 \\
    0 & 0 & d_{\rm grey}
    \end{bmatrix}
    \label{eq:8}
\end{equation}
where $d_{\rm grey}$ is the mean diffusivity of tumour cells in grey matter.

The white matter tumour cell tensor field $\boldsymbol{D}_{\rm white}(\boldsymbol{r})$ was built from the DTI-derived water diffusion tensor $\boldsymbol{D}_{\rm water}(\boldsymbol{r})$ using the method proposed by Jbabdi and colleagues in \cite{jbabdi_2005}. This step is motivated by the observation that, for mechanical \cite{johnson_2009} and molecular \cite{de_vleeschouwer_2017} reasons, tumour cells preferentially migrate along rather than across brain fibres, similarly to diffusing water molecules. The method consists in increasing the degree of anisotropy of the water diffusion tensor and scaling its mean diffusivity $\rm{MD}=\Tr\left(\boldsymbol{D}_{\rm water}\right)/3$ to account for the difference in diffusive behaviour between tumour cells and water molecules \cite{jbabdi_2005}:
\begin{equation}
    \boldsymbol{D}_{\rm white}(\boldsymbol{r}) = \frac{3 \, d_{\rm white}}{\sum_i \tilde{\lambda}_i(a)} \sum_i \tilde{\lambda}_i(a) \, \boldsymbol{e}_i(\boldsymbol{r}) \, \boldsymbol{e}_i^\top(\boldsymbol{r})
    \label{eq:9}
\end{equation}
where $d_{\rm white}$ is the mean diffusivity of tumour cells in white matter, $\boldsymbol{e}_i(\boldsymbol{r})$ is the $i$\textsuperscript{th} eigenvector of $\boldsymbol{D}_{\rm water}(\boldsymbol{r})$, and $\tilde{\lambda}_i(a) = l_i(a)\lambda_i$, with:
\begin{gather}
    \begin{bmatrix}
        l_1(a) \\
        l_2(a) \\
        l_3(a)
    \end{bmatrix} =
    \begin{bmatrix}
        a & a & 1 \\
        1 & a & 1 \\
        1 & 1 & 1
    \end{bmatrix}
    \begin{bmatrix}
        c_l \\
        c_p \\
        c_s
    \end{bmatrix},
    \label{eq:10} \\[4ex]
    c_l = \frac{\lambda_1-\lambda_2}{\lambda_1+\lambda_2+\lambda_3}, \quad c_p = \frac{2 \, (\lambda_2-\lambda_3)}{\lambda_1+\lambda_2+\lambda_3}, \quad c_s = \frac{3 \, \lambda_3}{\lambda_1+\lambda_2+\lambda_3}
    \label{eq:11}
\end{gather}
where $\lambda_i$ is the $i$\textsuperscript{th} eigenvalue of $\boldsymbol{D}_{\rm water}(\boldsymbol{r})$, $a \geq 1$ is a multiplicative factor introduced to increase the anisotropy of the tensor, and $c_l$, $c_p$, and $c_s$ are respectively the linear, planar, and spherical anisotropy measures of $\boldsymbol{D}_{\rm water}(\boldsymbol{r})$.

\begin{figure}[ht!]
    \centering
    \includegraphics[width=\textwidth]{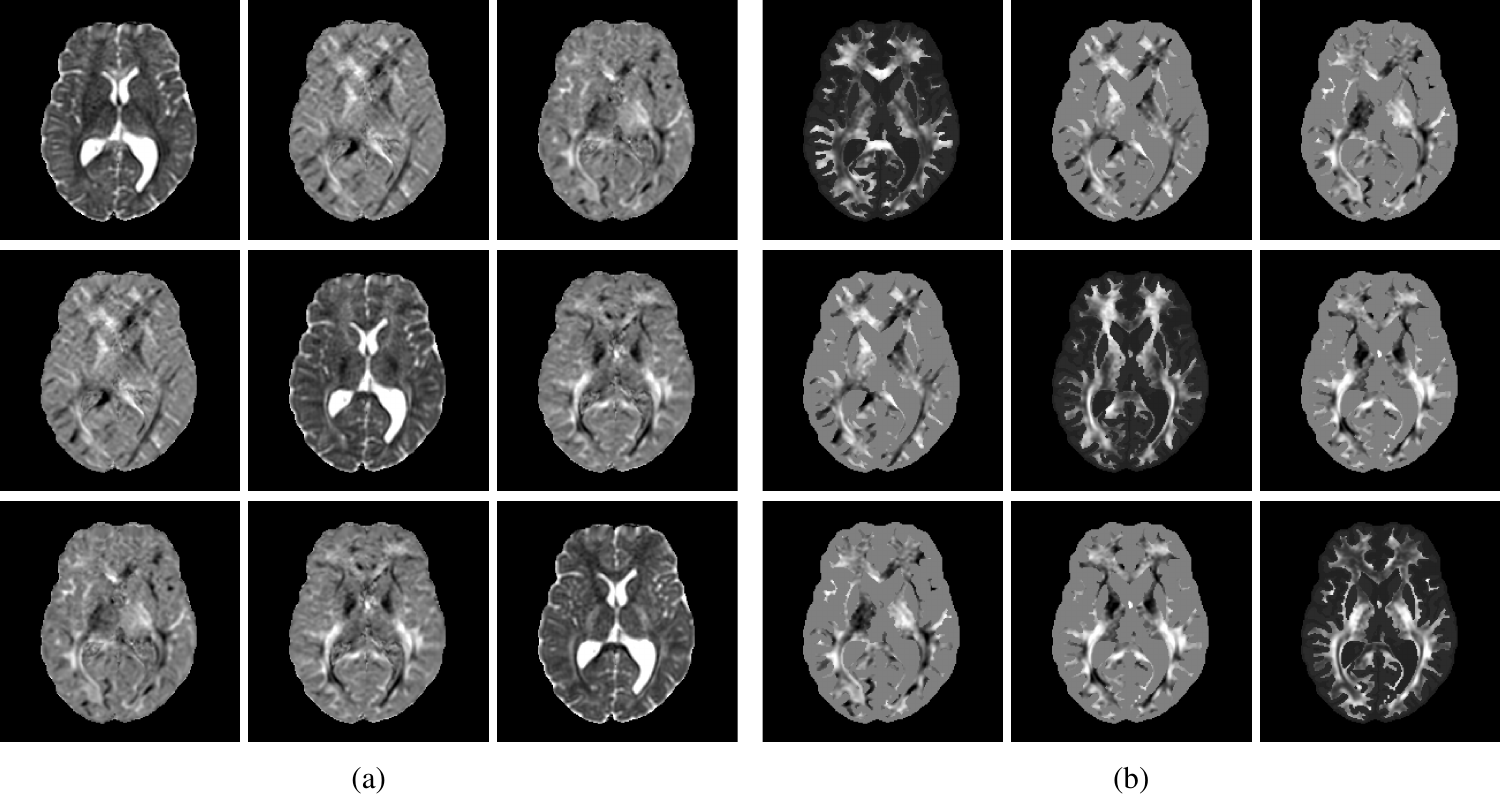}
    \caption{Example of processed DTI data. \textbf{(a)} DTI-derived water diffusion tensor field after susceptibility-induced distortion, eddy currents, and patient motion correction using FSL. \textbf{(b)} Tumour diffusion tensor field with increased anisotropy in white matter ($a=10$) and scaled diffusivity ($d_{\rm grey}/d_{\rm white}=0.1$) built from the water diffusion tensor field in panel (a) and the brain map in \Cref{fig:1}(c) as described in \Cref{subsubsec:2_3_6}. The subpanel located at row $i$ and column $j$ of panels (a) and (b) corresponds to the tensor component $d_{i, j}$.}
    \label{fig:2}
\end{figure}

For reasons that will become clearer in \Cref{subsec:2_4}, a unit tumour cell diffusion tensor was built at this stage by fixing $d_{\rm white}$ to 1.0. A value of 0.1 was chosen for the $d_{\rm grey}/d_{\rm white}$ ratio, as proposed previously in  \cite{swanson_2000} to account for the restricted migration of tumour cells in grey compared to white matter. This ratio was supposed constant among all subjects as it is expected to depend exclusively on the structural organisation of healthy white versus grey matter and not on the tumour characteristics. Similarly, the anisotropy factor $a$ was fixed to 10 for all subjects, as suggested in \cite{jbabdi_2005}. An example of processed DTI data is depicted in \Cref{fig:2}. The processed MR data of the 6 volunteers used in this study are publicly available at \url{https://lisaserver.ulb.ac.be/owncloud/index.php/s/KwEPG65gh1U7xNM}. Further details on these data are available in \Cref{app:b}.

\subsection{Dataset Synthesis}
\label{subsec:2_4}
A synthetic tumour dataset was generated from the processed MR data described hereabove. 200 fictitious tumours were grown over the segmented brain domain of each of the 6 volunteers from randomly picked seeds and parameter values using the reaction-diffusion model in \Cref{eq:4,eq:5,eq:6}, totalling 1,200 synthetic tumours. For each tumour, the cell density distribution was sampled at four imaging time points $t_1$, $t_2$, $t_3$, and $t_4$. 

Each tumour seed consisted of a $\SI{3}{\voxel} \times \SI{3}{\voxel} \times \SI{3}{\voxel}$ neighbourhood chosen among all segmented white matter voxels whose initial (i.e. at time $t=t_0$) normalised tumour cell density $c$ was set to 1. For each simulated tumour, an infiltration depth $\lambda = \sqrt{d_{\rm white}/\rho}$, a tumour front propagation velocity $v = 2 \sqrt{d_{\rm white} \, \rho}$, and two imaging time intervals $\Delta t_1$ and $\Delta t_2$ were randomly chosen from uniform distributions (floating point for $\lambda$ and $v$, integer for $\Delta t_{1,2}$). The value ranges of the uniform distributions are provided in \Cref{tab:1} and are of the same order of magnitude as in \cite{rockne_2010}.

\begin{table}[ht!]
    \centering
    \caption{Value ranges and units of the uniform distributions used to sample the tumour growth model parameters for the generation of the synthetic tumour dataset.}
    \begin{tabular}{cccc}
        \toprule
        & \textbf{Min} & \textbf{Max} & \textbf{Units} \\
        \midrule
        $\lambda$ & 0.5 & 2.0 & \si{\milli \meter} \\
        $v$ & 28.28 & 48.99 & \si{\milli \meter \per \year} \\
        $\Delta t_1$ & 90 & 180 & \si{\day} \\
        $\Delta t_2$ & 90 & 180 & \si{\day} \\
        \bottomrule
    \end{tabular}
    \label{tab:1}
\end{table}

Two additional time intervals, $\Delta t_3$ and $\Delta t_4$, were fixed to \SI{90}{\day} for validation purpose (see \Cref{subsec:2_6}). Starting from tumour emergence time $t_0$, the time intervals $\Delta t_{1-4}$ univocally define the four imaging time points $t_i=t_0+\sum_{j=1}^{i} \Delta t_j, \quad i=1, \ldots, 4$.

\begin{figure}[ht!]
    \centering
    \includegraphics[width=0.75\textwidth]{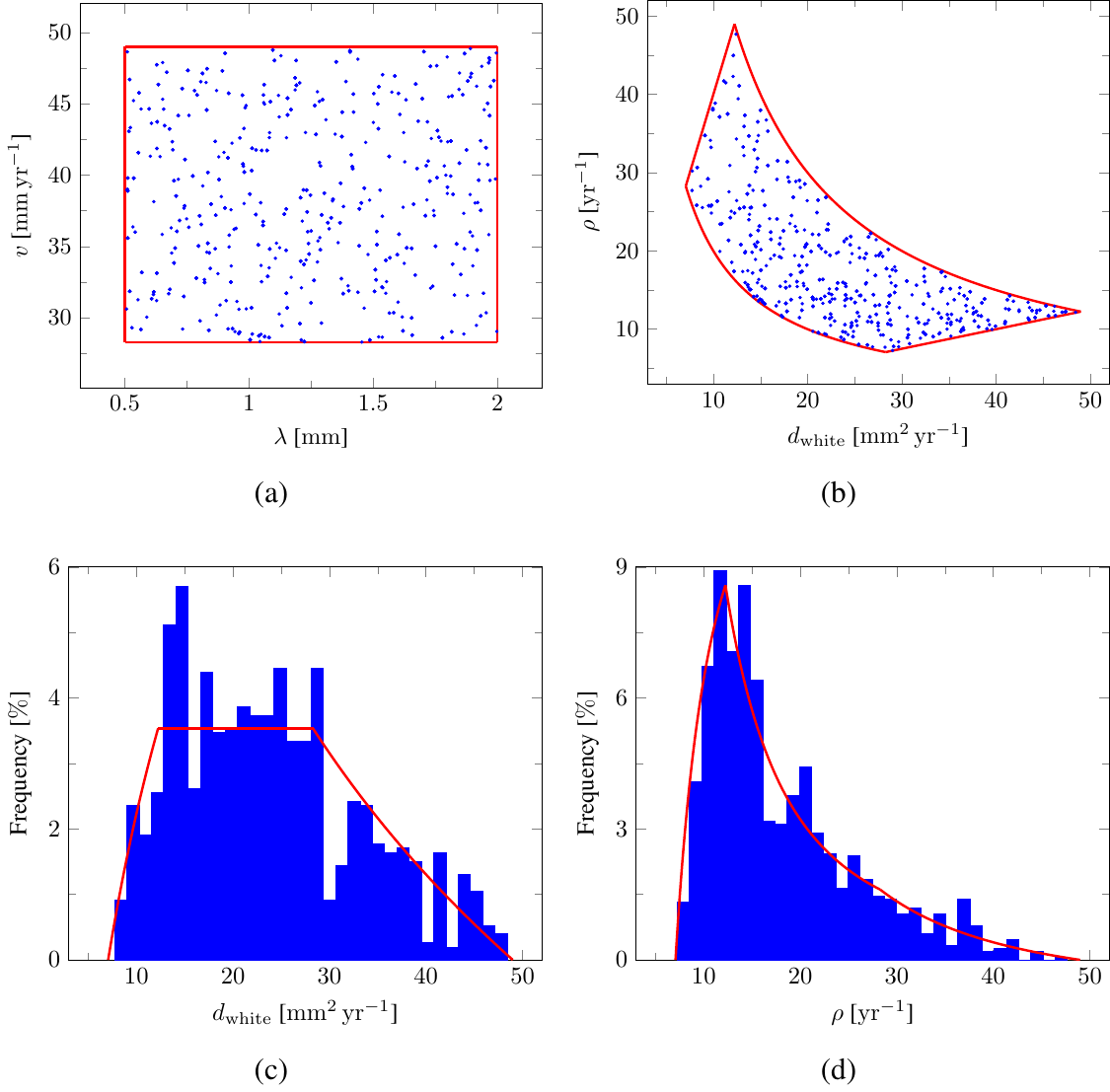}
    \caption{Sampling of the model parameters. \textbf{(a)} Empirical joint distribution of the $(\lambda, v)$ values sampled from uniform distributions (blue marks) with superimposed sampling domain boundary (red segments). \textbf{(b)} Corresponding joint distribution of the derived $(d_{\rm white}, \rho)$ values using \Cref{eq:12,eq:13} (blue marks) with superimposed sampling domain boundary (red curves). \textbf{(c)} Empirical marginal distribution of the derived $d_{\rm white}$ values (blue bars) with superimposed theoretical distribution (red curves). \textbf{(d)} Empirical marginal distribution of the derived $\rho$ values (blue bars) with superimposed theoretical distribution (red curves).}
    \label{fig:3}
\end{figure}

For each sampled couple of $(\lambda, v)$ values, a white matter diffusion rate value $d_{\rm white}$ and a proliferation rate value $\rho$ were derived as:
\begin{align}
    d_{\rm white} &= \frac{\lambda \, v}{2}
    \label{eq:12} \\
    \rho &= \frac{v}{2 \, \lambda}
    \label{eq:13} 
\end{align}
In this manner, a wide diversity of tumours can be uniformly sampled within a realistic range of infiltration depths and propagation velocities. Independently sampling $d_{\rm white}$ and $\rho$ values from uniform distributions may instead have resulted in too small (i.e. with too small values of $d_{\rm white}$ and $\rho$) or too large (i.e. with too large values of $d_{\rm white}$ and $\rho$) tumours. The empirical joint distribution of the sampled $(\lambda, v)$ values as well as the corresponding joint distribution of $(d_{\rm white}, \rho)$ and marginal distributions of $d_{\rm white}$ and $\rho$ are depicted in \Cref{fig:3}.

\begin{figure}[ht!]
    \centering
    \includegraphics[width=\textwidth]{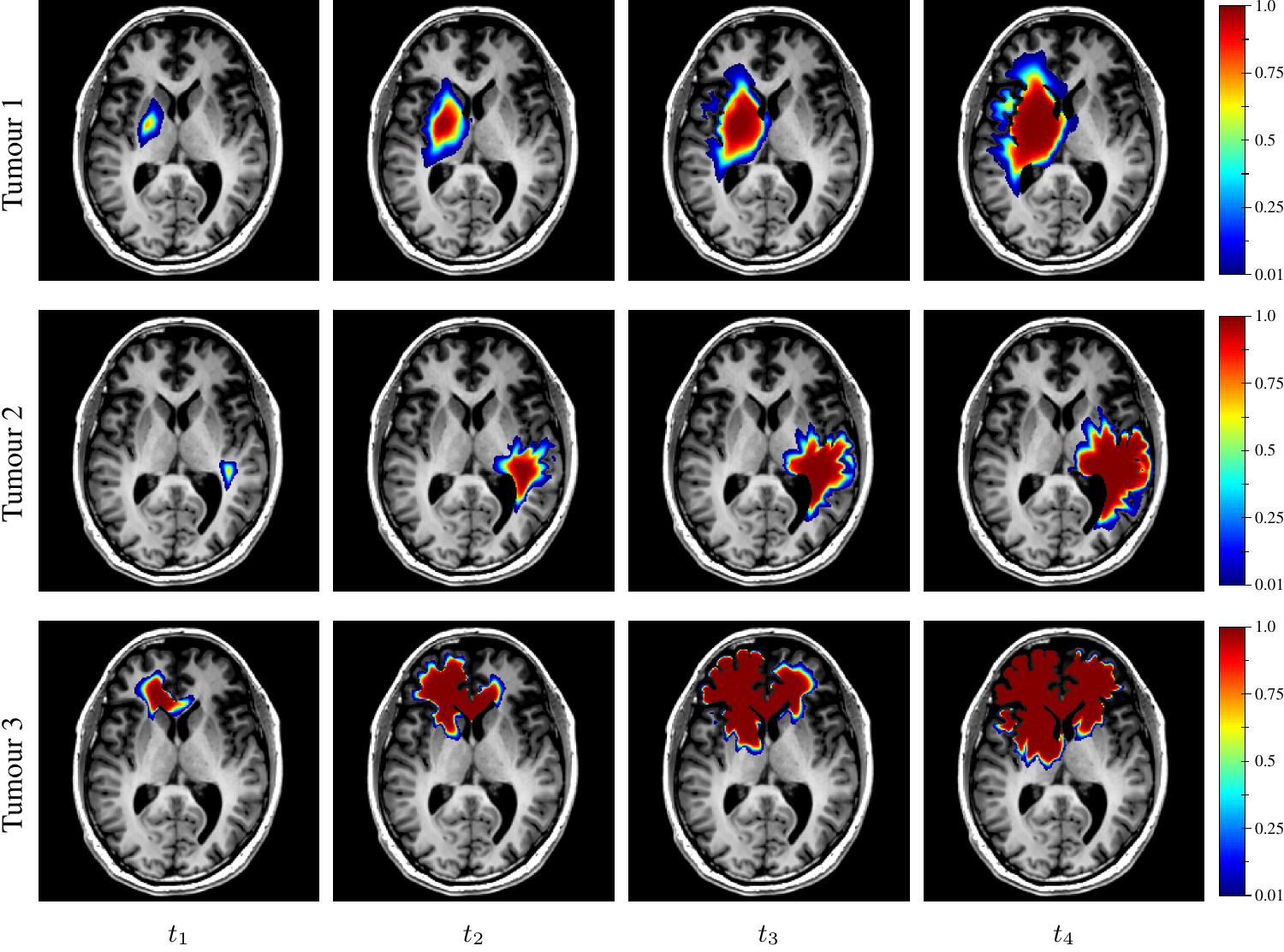}
    \caption{Examples of simulated tumour cell density distributions at times $t_{1-4}$ (1\textsuperscript{st} to 4\textsuperscript{th} columns, axial slices) from the MR data of the same subject as in \Cref{fig:1,fig:2}. The corresponding model parameter values are provided in \Cref{tab:2}.}
    \label{fig:4}
\end{figure}

For each synthetic tumour, a tumour cell diffusion tensor field $\boldsymbol{D}(\boldsymbol{r})$ was then obtained by multiplying the unit (unscaled) diffusion tensor derived as described in \Cref{subsubsec:2_3_6} by the derived value of $d_{\rm white}$. As a remainder, the ratio $d_{\rm grey}/d_{\rm white}$ was considered constant among tumours in this work (see \Cref{subsubsec:2_3_6}). A tumour was finally grown from the sampled seed, tumour cell diffusion tensor field $\boldsymbol{D}(\boldsymbol{r})$, and proliferation rate $\rho$ using the model and the simulated tumour cell distributions at times $t_{1-4}$ were stored. Examples of synthetic tumours are depicted in \Cref{fig:4}. The corresponding model parameter values are provided in \Cref{tab:2}.

\begin{table}[ht!]
    \centering
    \caption{Parameter values used for the tumour simulations in \Cref{fig:4}.}
    \begin{tabular}{ccccccccc}
        \toprule
        & $\boldsymbol{d_{\rm white}}$ \textbf{[\si{\square \milli \meter \per \year}]} & $\boldsymbol{\rho}$ \textbf{[\si{\per \year}]} & $\boldsymbol{\lambda}$ \textbf{[\si{\milli \meter}]} & $\boldsymbol{v}$ \textbf{[\si{\milli \meter \per \year}]} & $\boldsymbol{t_1}$ \textbf{[\si{\day}]} & $\boldsymbol{t_2}$ \textbf{[\si{\day}]} & $\boldsymbol{t_3}$ \textbf{[\si{\day}]} & $\boldsymbol{t_4}$ \textbf{[\si{\day}]} \\
        \midrule
        Tumour 1 & 10.87 & 31.77 & 1.71 & 37.16 & 175 & 328 & 418 & 508 \\
        Tumour 2 & 15.07 & 13.95 & 0.96 & 29.00 & 146 & 316 & 406 & 496 \\
        Tumour 3 & 41.49 & 11.31 & 0.52 & 43.33 & 137 & 242 & 332 & 422 \\
        \bottomrule
    \end{tabular}
    \label{tab:2}
\end{table}

The model was solved by a forward Euler finite difference approach with a time step $\Delta t=\SI{0.5}{\day}$ using a GPU implementation of the 3D standard stencil referenced in \cite{mosayebi_2010} for the computation of the divergence term in \Cref{eq:4}. The processing time was around \SI{1}{\milli \second} per iteration on a GeForce GTX 1080 GPU (NVIDIA, USA), leading to total simulation times in range \SIrange{0.72}{1.08}{\second} per tumour. All simulations were performed using a Python wrapping of our Tumour Growth Simulation ToolKit (TGSTK) written in C++/CUDA language based on VTK \cite{schroeder_2010} and publicly available at \url{https://github.com/cormarte/tgstk}. The toolkit documentation can be found at \url{https://cormarte.github.io/tgstk/html}.

\subsection{Deep Convolutional Neural Networks}
Two deep convolutional neural networks were implemented and trained to respectively address the cell density estimation and model parameter estimation problems introduced hereabove. All training steps were performed using the TensorFlow framework (version 2.5.0) \cite{abadi_2015} in Python on a GeForce RTX 3090 GPU (NVIDIA, USA).

\subsubsection{Cell Density Estimation}
The first problem addressed was to reconstruct a tumour cell density distribution from (i) two imaging contours---$\Gamma_1$ and $\Gamma_2$---obtained through an imaging process described by \Cref{eq:2} for two detectability threshold values $c_1$ and $c_2$ at a given imaging time and (ii) a unit (unscaled) tumour cell diffusion tensor field derived from DTI data as presented in \Cref{subsubsec:2_3_6}. 

This approach is motivated by the properties of the asymptotic travelling wave solution admitted by \Cref{eq:1} for constant coefficients and on the infinite cylinder, whose profile decreases exponentially with decay constant $\lambda$ (see \Cref{sec:1}). For such a solution, the value of $\lambda$ can be trivially estimated given the distance between two cell density iso-contours, and an approximate tumour cell density distribution can subsequently be reconstructed for sufficiently large $x$. Here, we assess the ability of deep neural networks to build an approximate solution of \Cref{eq:4,eq:5,eq:6} in the more general case of a complex domain and variable anisotropic diffusion tensor field, and from 2 imaging contours obtained by \Cref{eq:2}---which do not necessarily coincide with cell density iso-contours as discussed previously \cite{martens_2021}.

In a clinical setting, the value of the white matter diffusion rate $d_{\rm white}$ used to scale the tumour cell diffusion tensor in \Cref{eq:9} is unknown---and its assessment will be addressed in the next section. Therefore this information is not considered for this problem. In contrast, the preferred migration directions of tumour cells along the white matter tracts can be assessed from clinical DTI acquisitions as described in \Cref{subsubsec:2_3_6} and may be used for the estimation of the tumour cell density distribution. This motivates the introduction of the unit unscaled diffusion tensor field as an input of this problem, in addition to the $\Gamma_1$ and $\Gamma_2$ contours.

\begin{figure}[ht!]
    \centering
    \includegraphics[width=\textwidth]{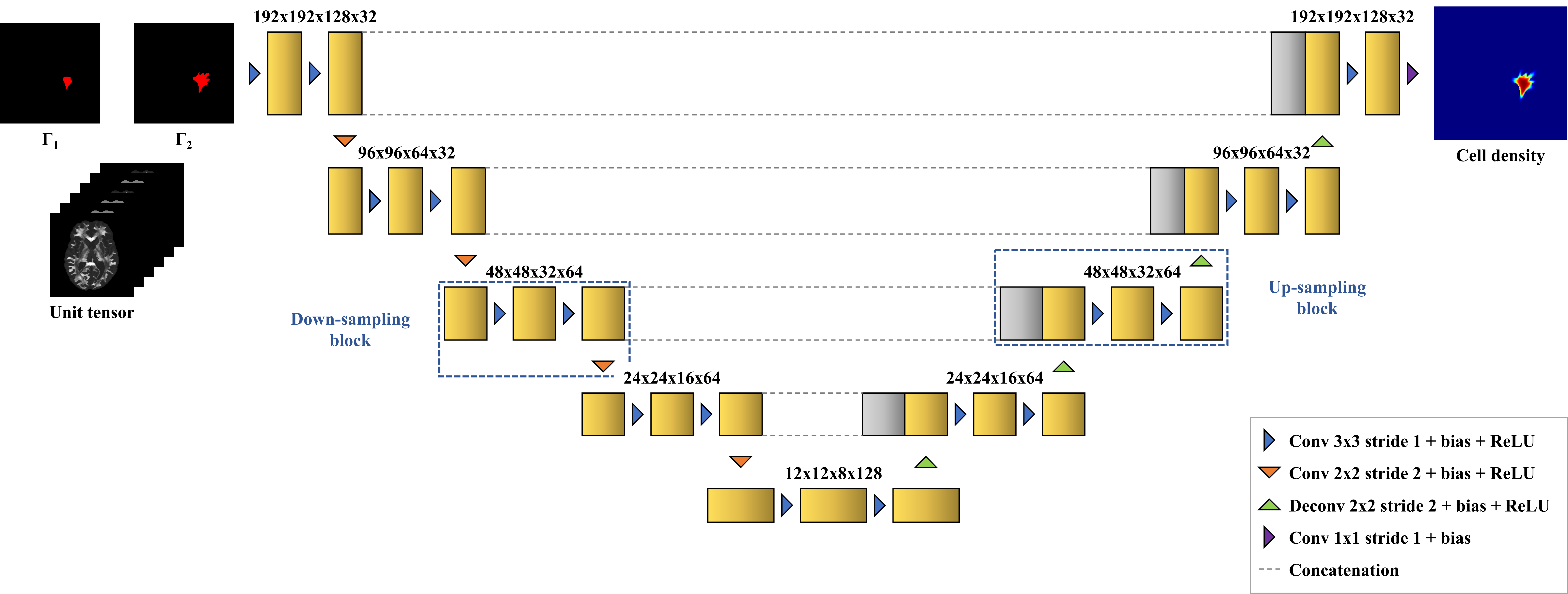}
    \caption{3D U-Net architecture \cite{ronneberger_2015} with its feature map dimensions used for cell density estimation. The network takes as input volumes of dimensions $192 \times 192 \times 128$ with 8 channels corresponding to the 2 contours $\Gamma_1$ and $\Gamma_2$ and the 6 independent components of the unit (unscaled) tumour cell diffusion tensor field and outputs a cell density map with the same spatial dimensions.}
    \label{fig:5}
\end{figure}

To address this problem, a 3D DCNN based on the U-Net architecture \cite{ronneberger_2015} was implemented, as it has been successfully applied to many medical imaging problems previously \cite{altaf_2019,siddique_2021}. The network consists of 4 down-sampling blocks, 4 up-sampling blocks, and 1 output block. Each down-sampling block is made of 2 convolutional layers with kernel size $3 \times 3$ and stride 1, followed by a bias-adding layer and a rectified linear unit (ReLU) activation layer. A convolutional layer with kernel size $2 \times 2$ and stride 2 is added at the end of the block to reduce the feature map dimensions by a factor 2. The up-sampling blocks are identical to the down-sampling blocks except that the last convolutional layer is replaced by a transposed convolution layer with kernel size $2 \times 2$ and stride 2, followed by a bias-adding layer and a ReLU activation layer to expand the feature maps dimensions by a factor 2. Skip connections are added between the output of the second ReLU activation layer of each down-sampling block and the input of the corresponding upsampling block with the same spatial dimensions, implemented as a concatenation operation. The output block has the same structure as the down-sampling blocks except that the last convolutional layer is replaced by a convolutional layer with kernel size $1 \times 1$ and stride 1 followed by a bias-adding layer but no activation layer to merge the last 32 feature maps into a single tumour cell density map. The network architecture with its feature map dimensions is depicted in \Cref{fig:5}.

The network takes as input tensors of shape $192 \times 192 \times 128 \times 8$ (width $\times$ height $\times$ depth $\times$ channels). The first 2 channels are fed with the two binary regions respectively delimited by $\Gamma_1$ and $\Gamma_2$. These regions were obtained by thresholding each generated tumour cell distribution at the second imaging time point $t_2$ with threshold values $c_1$ and $c_2$ of 0.80 and 0.16, as previously proposed in \cite{swanson_2008} for the enhancing core and oedema outlines, respectively. The last 6 channels correspond to the 6 independent components of the unit (unscaled) tumour cell diffusion tensor (see \Cref{subsubsec:2_3_6}). 

To evaluate the generalisation ability of the model, the dataset was further split into training and test sets in proportion \SI{83}{\percent}--\SI{17}{\percent} on a patient basis (i.e. the 200 tumours generated from the MR data of the last patient are kept for evaluation purpose). The network was trained using the Adam optimiser \cite{kingma_2014} (learning rate: \num{e-4}, $\beta_1$: 0.9, $\beta_2$: 0.999, $\epsilon$: \num{e-6}) and the mean absolute error (MAE) loss over mini-batches of size 1. Data augmentation was performed by applying random shifts in range $\pm \SI{15}{\voxel}$ in the three spatial dimensions to each batch. Rotations were not applied as they would imply transformation of the tensor components, resulting in longer execution times for on-the-fly augmentation or larger dataset size for offline augmentation. The training was stopped early after no improvement in the test loss for 100 epochs. The network parameter values that provided the best test loss value (MAE = \num{1.24e-4}) were kept, which occurred after 876 epochs.

\subsubsection{Parameter Estimation}
The second problem addressed is to estimate the value of the model parameters $d_{\rm white}$ and $\rho$ (or equivalently of the derived parameters $\lambda = \sqrt{d_{\rm white}/\rho}$ and $v = 2 \sqrt{d_{\rm white} \, \rho}$) from (i) three imaging contours: two imaging contours---$\Gamma_1$ and $\Gamma_2$---obtained by \Cref{eq:2} for two different threshold values $c_1$ and $c_2$ at imaging time $t_2$ and a third imaging contour $\Gamma_3$ obtained for the same $c_2$ threshold value at the earlier imaging time $t_1$ and (ii) the unit (unscaled) tumour cell diffusion tensor field. The time interval $\Delta t_2$ between $t_1$ and $t_2$ is also considered as an input of the problem. 

As for the cell density estimation problem, the motivation for such inputs lies in the properties of the asymptotic travelling wave solution of \Cref{eq:1} (see \Cref{sec:1}), whose profile decay constant $\lambda=\sqrt{d/\rho}$ can be assessed from two cell density iso-contours at a given time point as mentioned hereabove. In addition, the propagation velocity of the tumour front $v=2\sqrt{d \, \rho}$ can similarly be assessed from the distance between a same cell density iso-contour taken at two different time points given their temporal spacing. The knowledge of $\lambda$ and $v$ can finally be used to assess the individual values of $d$ and $\rho$. Here again, we assess the ability of deep neural networks to generalise these properties in the case of a complex domain and variable anisotropic diffusion tensor field from 3 threshold-like imaging contours obtained by \Cref{eq:2}. The same remark as for the previous section holds regarding the possibility of deriving a unit tumour diffusion tensor from DTI data, which can provide additional information for the estimation of the steepness and velocity of the tumour front.

The DCNN implemented for this task is a convolutional encoder. The network consists of 6 convolutional down-sampling blocks and a fully connected output block. Each down-sampling block is made of 2 convolutional layers with kernel size $3 \times 3$ and stride 1, followed by a bias-adding layer and a ReLU activation layer. A convolutional layer with kernel size $2 \times 2$ and stride 2 is added at the end of the block to reduce the feature map dimensions by a factor 2. The output block flattens the $3 \times 3 \times 2 \times 8$ output of the last down-sampling block and concatenates a $1 \times 1$ (width $\times$ channels) tensor to the flattened vector to feed the imaging time interval $\Delta t_2$. A fully connected layer followed by a bias-adding layer but no activation layer is finally used to merge the last 145 components into 2 scalar values for $\lambda$ and $v$.

The network takes as input tensors of shape $192 \times 192 \times 128 \times 9$ (width $\times$ height $\times$ depth $\times$ channels). The first 3 channels are fed with the binary regions respectively delimited by $\Gamma_1$, $\Gamma_2$, and $\Gamma_3$. These regions were respectively obtained by thresholding each generated tumour cell distribution at time $t_2$ with threshold values of 0.80 ($\Gamma_1$) and 0.16 ($\Gamma_2$) and the distribution at time $t_1$ with a value threshold value of 0.16 ($\Gamma_3$) \cite{swanson_2008}. The last 6 channels correspond to the 6 independent components of the unit (unscaled) tumour cell diffusion tensor (see \Cref{subsubsec:2_3_6}). To account for their different value range and scale, the target values of $\lambda$ and $v$ were standardised using the theoretical mean and variance of the respective uniform distributions from which they were sampled.

The same training/test splitting as for the tumour cell density estimation network was applied to the dataset. The network was trained using the Adam optimiser \cite{kingma_2014} (learning rate: \num{e-4}, $\beta_1$: 0.9, $\beta_2$: 0.999, $\epsilon$: \num{e-6}) and the mean squared error (MSE) loss. Data augmentation was performed by applying random shifts in range $\pm \SI{15}{\voxel}$ in the three spatial dimensions to each input batch. Early stopping was applied if no improvement was observed in the test loss for 100 epochs. The network parameter values that provided the best test loss value (MSE = \num{6.75e-2}) were kept, which occurred after 628 epochs.

\begin{figure}[ht!]
    \centering
    \includegraphics[width=0.8373\textwidth]{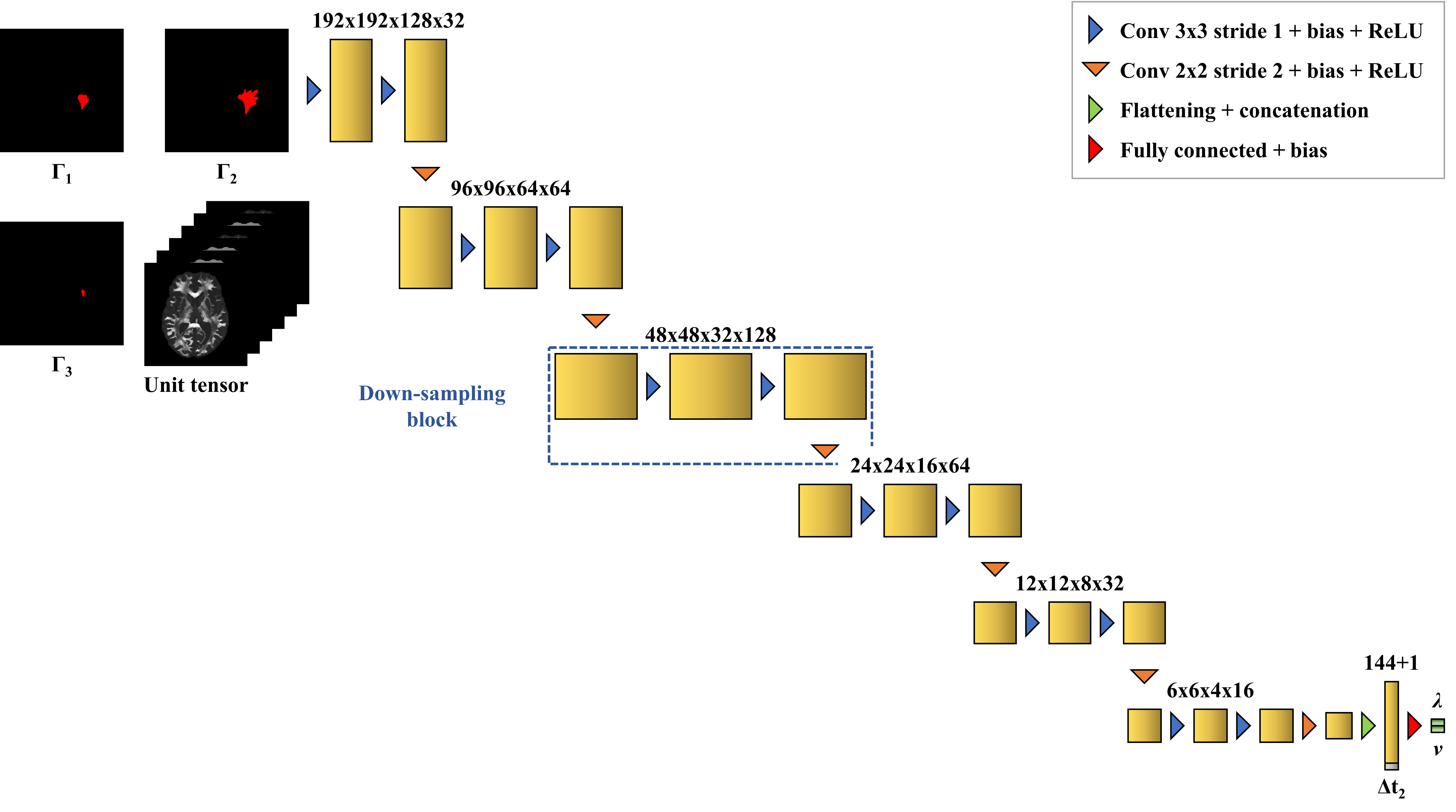}
    \caption{3D convolutional regressor architecture with its feature map dimensions used for parameter estimation. The network takes as input volumes of dimensions $192 \times 192 \times 128$ with 9 channels corresponding to the 3 contours $\Gamma_1$, $\Gamma_2$, and $\Gamma_3$ and the 6 independent components of the unit (unscaled) tumour cell diffusion tensor field as well as the time interval $\delta t_2$ between $\Gamma_3$ and $\Gamma_2$, and outputs estimated values of $\lambda$ and $v$.}
    \label{fig:6}
\end{figure}

\subsection{Validation}
\label{subsec:2_6}
To validate and illustrate our approach, we conducted the following numerical experiment: Starting from the tumour cell density distribution estimated at time $t_2$ from $\Gamma_1$ and $\Gamma_2$ using our first network (\Cref{fig:5}), as well as the values of $d_{\rm white}$ and $\rho$ estimated from $\Gamma_1$, $\Gamma_2$, and $\Gamma_3$ using our second network (\Cref{fig:6}) and \Cref{eq:12,eq:13}, we computed a tumour cell density distribution using the reaction-diffusion model at times $t_3$ and $t_4$, \SI{90}{\day} and \SI{180}{\day} later, respectively. We then compared the estimated distributions to the actual tumour cell density distributions at times $t_3$ and $t_4$---i.e. these obtained for the true cell density distribution at time $t_2$ as well as the true values of $d_{\rm white}$ and $\rho$---using the MAE computed voxelwise within the $c > 0.01$ contour. This latter restriction prevents background or weakly invaded voxels to artificially lower the MAE. The Hausdorff distance $d_\text{Hausdorff}$ and the average symmetric surface distance (ASSD) $d_\text{ASSD}$ between the imaging contours obtained from the true and estimated tumour cell density distributions for threshold values of $c_1$ and $c_2$ were also computed for each test tumour and time point, as given by:
\begin{align}
d_\text{Hausdorff}(A, B) &= \max\left\{ \max_{b \in B} \left\{ \min_{a \in A} d(a, b) \right\}, \max_{a \in A} \left\{ \min_{b \in B} d(a, b) \right\}\right\},
\label{eq:14} \\
d_\text{ASSD}(A, B) &= \frac{1}{\lvert A \rvert + \lvert B \rvert}\left(\sum_{b \in B} \min_{a \in A} d(a, b) + \sum_{a \in A} \min_{b \in B} d(a, b)\right),
\label{eq:15}
\end{align}
where $d(x,y)$ is the Euclidian distance between elements $x$ and $y$, and $\lvert X \rvert$ is the cardinal of a set $X$.

It should be noted that minor post-processing was applied to the estimated tumour cell distributions at time $t_2$ provided by the first network prior to the computation of the densities at times $t_3$ and $t_4$. First, the cell density of non-brain voxels (i.e. cerebrospinal fluid and background voxels) was set to 0 as small ($\sim \num[retain-unity-mantissa=false]{1e-5}$) but non-zero values were observed for some of these voxels in the predicted tumour cell distribution. Second, maximum densities were clipped to~1 as small overshootings were also occasionally observed. Third, voxels located outside the largest connected region with densities above \num{1e-6} were also set to 0 since small local maxima ($\sim \num[retain-unity-mantissa=false]{1e-5}$) were sporadically observed far from the tumour core, which gave rise to new tumour foci throughout the simulation. These post-processing steps allow to correct for inaccuracies in the non-constrained output of our convolutional network and ensure numerical stability of the reaction-diffusion model solution at later times.

Finally, to demonstrate the applicability of our approach in a clinical context, a cell density map was generated from the retrospective MR data of the GBM patient (see \Cref{subsec:2_2,subsec:2_3}). To this extend, the segmented enhancing core and oedema regions (see \Cref{subsubsec:2_3_5}) were provided to the first network along with the derived unit tumour cell diffusion tensor.

%A sensitivity analysis was then conducted on the tumour detectability threshold. To this end, additional contours were generated using varying threshold values in range $\pm \SI{10}{\percent}$ with step $\SI{1}{\percent}$ and fed to both networks.

\section{Results}
\begin{figure}[H]
    \centering
    \includegraphics[width=0.375\textwidth]{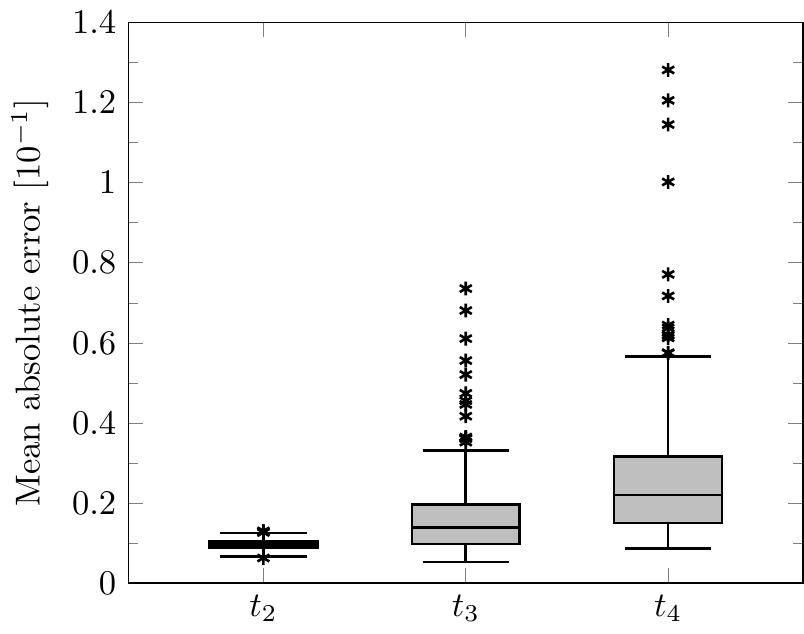}
    \caption{Boxplots of the mean absolute error distribution within the $c > 0.01$ contour computed voxelwise over the whole test set for times $t_2=\Delta t_1+\Delta t_2 \in \SIrange[range-phrase={,}\ , range-units=brackets, open-bracket=[, close-bracket=]]{180}{360}{\day}$ (see \Cref{tab:1}), $t_3=t_2+\SI{90}{\day}$, and $t_4=t_2+\SI{180}{\day}$. Horizontal line: median, box: interquartile range, whiskers: $\pm 1.5$ interquartile range, asterisks: outliers.}
    \label{fig:10}
\end{figure}

\begin{figure}[H]
    \centering
    \includegraphics[width=0.75\textwidth]{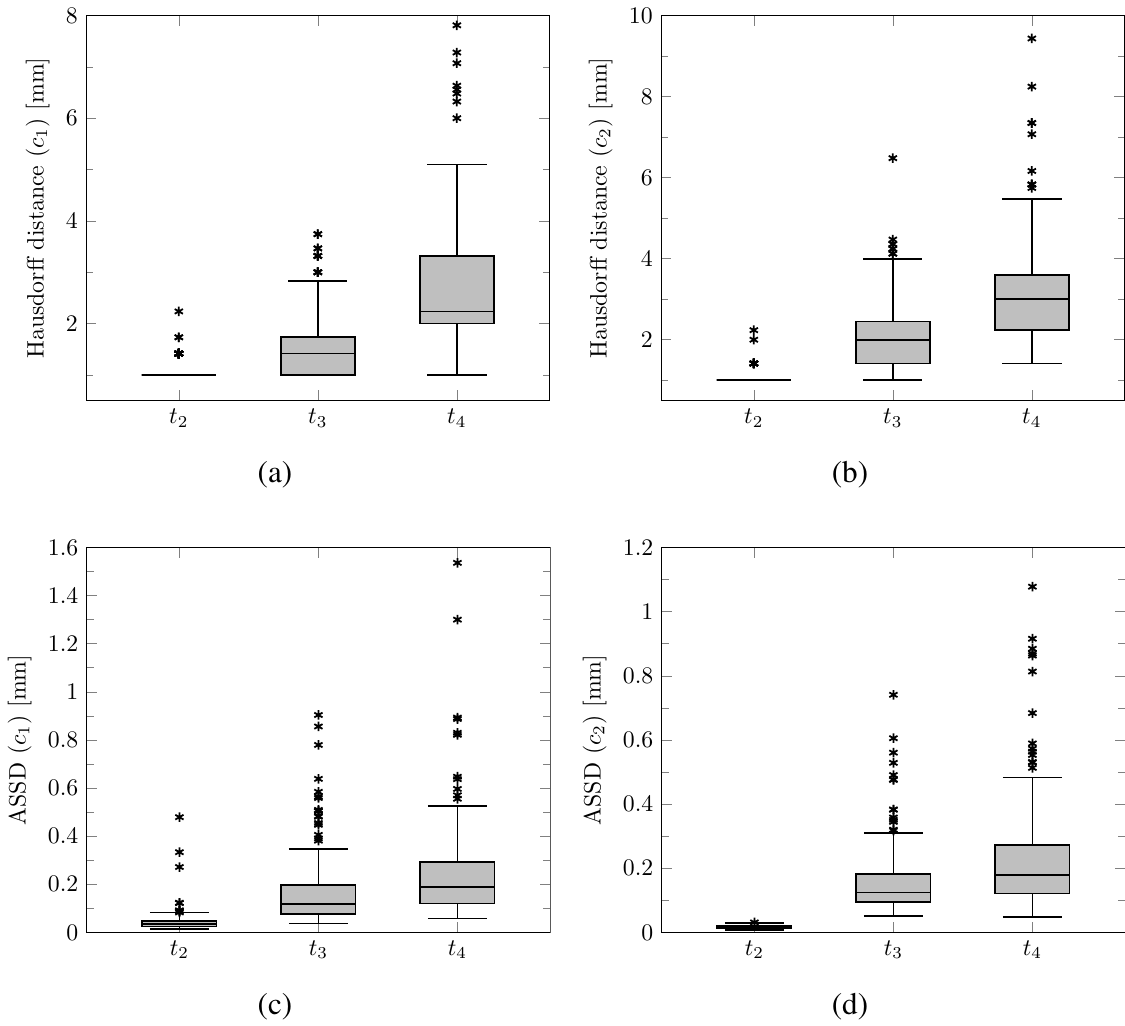}
    \caption{Boxplots of the Hausdorff distance and average symmetric surface distance (ASSD) distributions computed between the two imaging contours over the whole test set for times $t_2=\Delta t_1+\Delta t_2 \in \SIrange[range-phrase={,}\ , range-units=brackets, open-bracket=[, close-bracket=]]{180}{360}{\day}$ (see \Cref{tab:1}), $t_3=t_2+\SI{90}{\day}$, and $t_4=t_2+\SI{180}{\day}$. \textbf{(a)}--\textbf{(b)} Hausdorff distances computed between the true and estimated imaging contours obtained for threshold values of $c_1=0.80$ and $c_2=0.16$, respectively. \textbf{(c)}--\textbf{(d)} ASSD values computed between the true and estimated imaging contours obtained for threshold values of $c_1=0.80$ and $c_2=0.16$, respectively. Horizontal line: median, box: interquartile range, whiskers: $\pm 1.5$ interquartile range, asterisks: outliers.}
    \label{fig:11}
\end{figure}

\begin{figure}[H]
    \centering
    \includegraphics[width=0.75\textwidth]{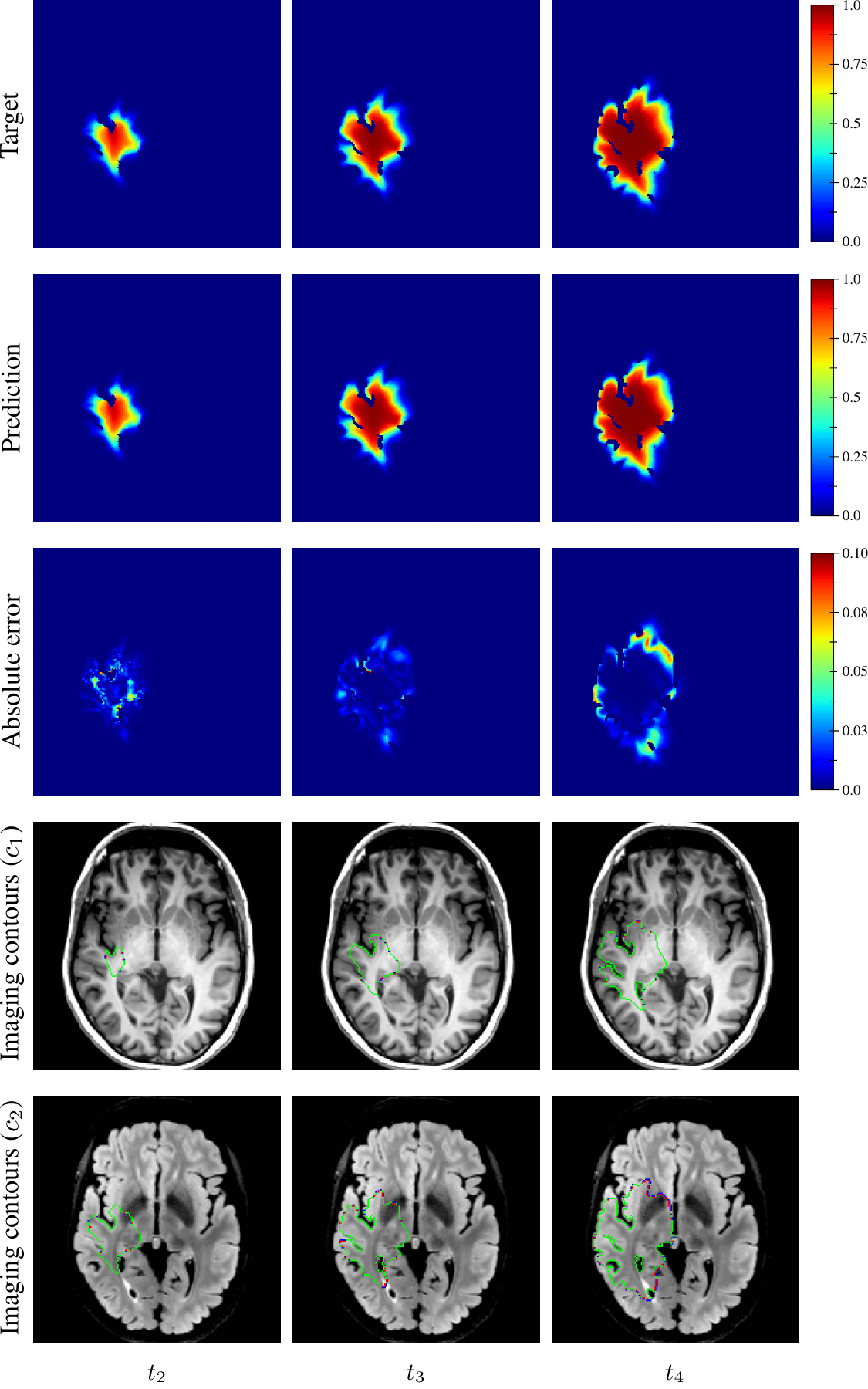}
    \caption{Example of true (1\textsuperscript{st} row) and estimated (2\textsuperscript{nd} row) tumour cell density distributions at times $t_{2-4}$ (1\textsuperscript{st} to 3\textsuperscript{rd} column, axial slices) along with the corresponding absolute error maps (3\textsuperscript{rd} row) for a test tumour ($d=\SI{43.47}{\square \milli \meter \per \year}$, $\rho=\SI{11.22}{\per \year}$, $t_1=\SI{94}{\day}$, $t_2=\SI{264}{\day}$). The imaging contours for threshold values $c_1=0.80$ and $c_2=0.16$ superimposed to the T1 and T2 FLAIR image are depicted in the 4\textsuperscript{th} and 5\textsuperscript{th} rows, respectively. The blue, red, and green segments respectively correspond to the target, prediction, and overlapping contour voxels. The predicted cell distribution at time $t_2$ was provided by the first network (\Cref{fig:5}). The estimated cell distribution at times $t_3$ and $t_4$ were computed using the reaction-diffusion model from the cell distribution predicted at time $t_2$ and the predicted model parameters values provided by the second network (\Cref{fig:6}).}
    \label{fig:9}
\end{figure}

The distribution of the mean absolute error computed over the test set between the true and estimated tumour cell distributions at time $t_2$ within the $c > 0.01$ contour is summarised by a boxplot in \Cref{fig:10} (1\textsuperscript{st} plot). The corresponding median value was \num{9.58e-3}. Boxplots of the Hausdorff distance and ASSD distributions computed over the test set between the true and estimated imaging contours at time $t_2$ for threshold values $c_1=0.80$ and $c_2=0.16$ are provided in \Cref{fig:11} (1\textsuperscript{st} plots). An example of true and estimated tumour cell density distributions at time $t_2$ from the test set is depicted in \Cref{fig:9} (1\textsuperscript{st} column), along with the corresponding absolute error map as well as the true and estimated imaging contours for threshold values $c_1=0.80$ and $c_2=0.16$. Additional examples are provided in \Cref{app:c}. All predicted tumour cell distributions at time $t_2$ used in \Cref{fig:9,fig:10,fig:11} were provided by the first network (\Cref{fig:5}).

\begin{figure}[H]
    \centering
    \includegraphics[width=0.75\textwidth]{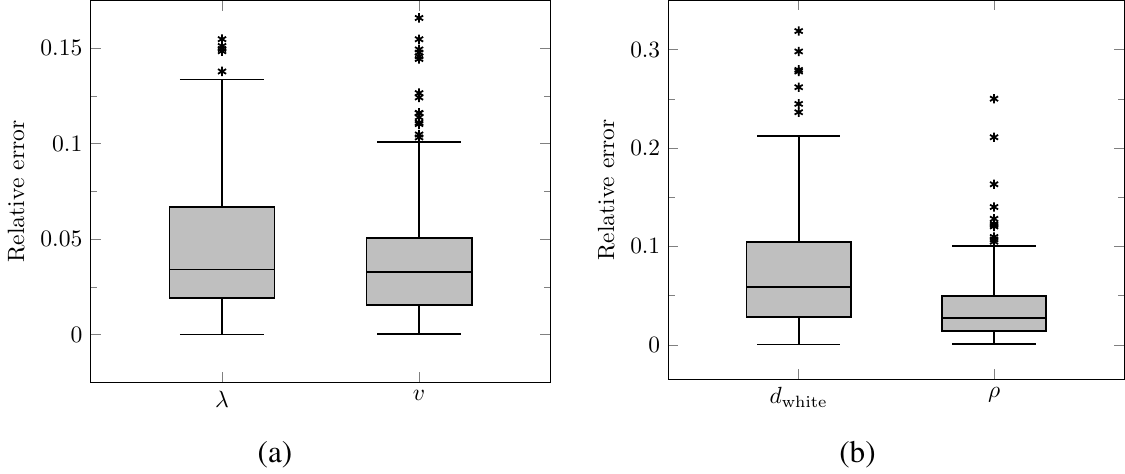}
    \caption{Boxplots of the relative error on the predicted model parameter values evaluated on the test set. \textbf{(a)} Relative errors on the estimated values of $\lambda$ and $v$ provided by the second network (\Cref{fig:6}). \textbf{(b)} Corresponding relative errors on the derived values of $d_{\rm white}$ and $\rho$ using \Cref{eq:12,eq:13}. Horizontal line: median, box: interquartile range, whiskers: $\pm 1.5$ interquartile range, asterisks: outliers.}
    \label{fig:7}
\end{figure}

\begin{figure}[H]
    \centering
    \includegraphics[width=0.75\textwidth]{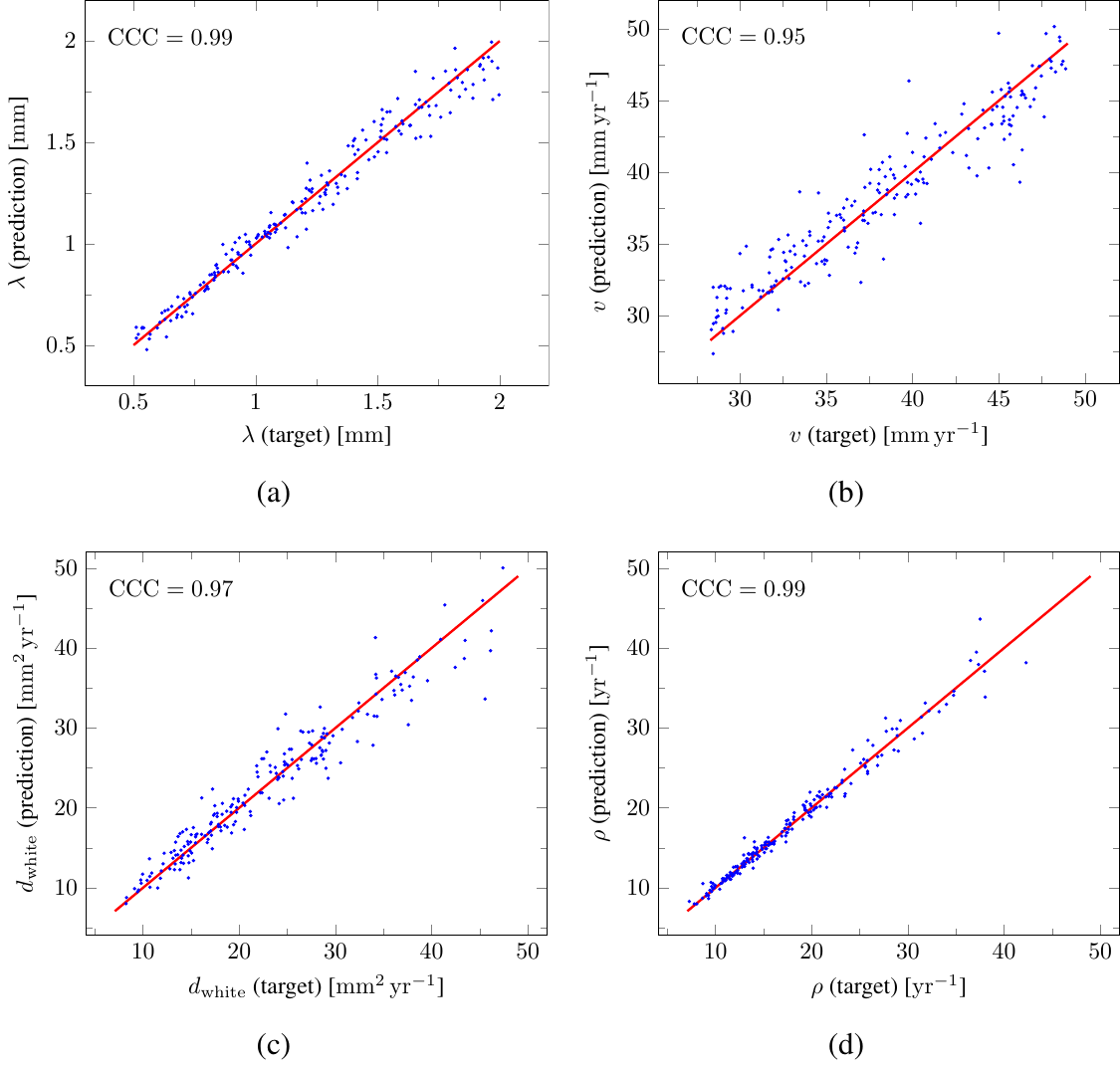}
    \caption{Scatterplots of the true versus predicted values of the model parameters from the test set. \textbf{(a)}--\textbf{(b)} True versus predicted values of $\lambda$ and $v$ provided by the second network (\Cref{fig:6}). \textbf{(c)}--\textbf{(d)} True versus estimated values of $d_{\rm white}$ and $\rho$ derived from the predicted values of $\lambda$ and $v$ using \Cref{eq:12,eq:13}. For each plot, the identity function is superimposed in red and the corresponding Lin's concordance correlation coefficient $\mathrm{CCC}$ is provided.}
    \label{fig:8}
\end{figure}

The distributions of the relative error on the values of $\lambda$ and $v$ computed at time $t_2$ over the test set as well as on the values of $d_{\rm white}$ and $\rho$ derived with \Cref{eq:12,eq:13} are summarised by boxplots in \Cref{fig:7}. The corresponding median relative errors were \SI{3.41}{\percent}, \SI{3.30}{\percent}, \SI{5.86}{\percent}, and \SI{2.75}{\percent} for $\lambda$, $v$, $d_{\rm white}$, and $\rho$, respectively. The predicted versus estimated values of $\lambda$ and $v$ as well as of $d_{\rm white}$ and $\rho$ from the test set are plotted in \Cref{fig:8}. The corresponding Lin's concordance correlation coefficients ($\mathrm{CCC}$) \cite{lin_1989} were 0.99, 0.95, 0.97, and 0.99 for $\lambda$, $v$, $d_{\rm white}$, and $\rho$, respectively.

As for imaging time $t_2$, the distributions of the mean absolute error computed over the test set between the true and estimated tumour cell distributions at times $t_3$ and $t_4$ within the $c > 0.01$ contour are summarised by boxplots in \Cref{fig:10} (2\textsuperscript{nd} and 3\textsuperscript{rd} plot, respectively). The corresponding median values were \num{1.38e-2} and \num{2.20e-2} for $t_3$ and $t_4$, respectively. Boxplots of the Hausdorff distance and ASSD distributions computed over the test set between the true and estimated imaging contours at times $t_3$ and $t_4$ for threshold values $c_1=0.80$ and $c_2=0.16$ are also provided in \Cref{fig:11} (2\textsuperscript{nd} and 3\textsuperscript{rd} plots). The true and estimated tumour cell density distributions at times $t_3$ and $t_4$ are depicted in \Cref{fig:9} (2\textsuperscript{nd} and 3\textsuperscript{rd} column, respectively) for the same test case as for time $t_2$, along with the corresponding absolute error maps as well as the true and estimated imaging contours for threshold values $c_1=0.80$ and $c_2=0.16$. Additional examples are provided in \Cref{app:c}. A loss of accuracy in the estimated tumour cell density distributions over simulated time is observed in \Cref{fig:10,fig:11,fig:9}. The estimated tumour cell density distributions at times $t_3$ and $t_4$ used in \Cref{fig:9,fig:10,fig:11} were computed using the reaction-diffusion model as described in \Cref{subsec:2_6} from (i) the cell density distribution predicted at time $t_2$ provided the first network (\Cref{fig:5}) and (ii) the predicted model parameters values provided by the second network (\Cref{fig:6}).

\begin{figure}[H]
    \centering
    \includegraphics[width=0.75\textwidth]{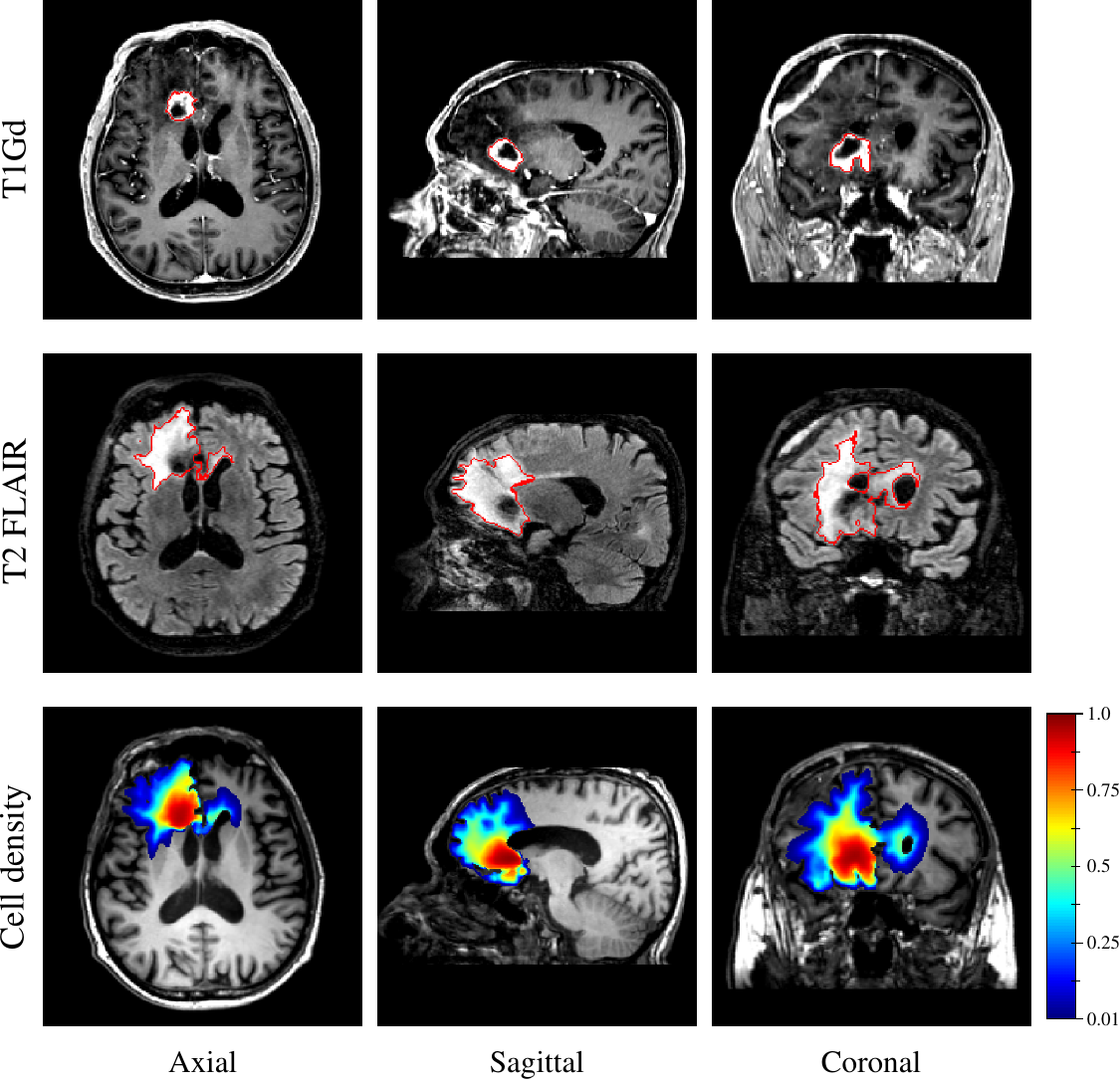}
    \caption{T1Gd image (1\textsuperscript{st} row), T2 FLAIR image (2\textsuperscript{nd} row), and estimated tumour cell density distribution using the first network (3\textsuperscript{rd} row) for an IDH-wildtype glioblastoma patient in axial (1\textsuperscript{st} column), sagittal (2\textsuperscript{nd} column), and coronal (3\textsuperscript{rd} column) planes. The contours of the segmented enhancing core and peritumour vasogenic oedema are superimposed in red on the T1Gd (1\textsuperscript{st} row) and T2 FLAIR (2\textsuperscript{nd} row) images, respectively.}
    \label{fig:12}
\end{figure}

Finally, the estimated tumour cell distribution for the studied GBM patient provided by the first network (see \Cref{fig:5}) is depicted in \Cref{fig:12} along with the T1Gd and T2 FLAIR images with superimposed segmented enhancing core and oedema contours, respectively.

\vfill
\clearpage

\section{Discussion}
Reaction-diffusion models have been studied for decades to capture the growth of gliomas but severe limitations regarding the estimation of their initial conditions and parameter values have restrained their use as a proper personalised clinical tool. In this work, we showed the ability of DCNNs to circumvent these limitations, opening a wide range of opportunities in the field. Our approach only requires to (i) derive a unit tensor field from clinical DTI data as described herein, accounting for the preferential migration of tumour cell along white matter tracts and (ii) extract three contours obtained through a cell density threshold-like imaging process described by \Cref{eq:2} for two different threshold values and time points. 

Regarding the second requirement, the outlines of the peritumour vasogenic oedema and enhancing core have been proposed previously \cite{swanson_2008}, respectively visible on T2 FLAIR and T1Gd MR images acquired in routine for glioma follow-up. Nevertheless, it is worth noticing that peritumour vasogenic oedema does not strictly speaking correspond to a region of tumour cell invasion but results from an accumulation of extracellular fluid originating from tumour-induced alterations of the blood-brain barrier \cite{hawkins-daarud_2013, lin_2013} and changes in hydrodynamic pressure \cite{lu_2004}. Consequently, the T2 FLAIR imaging process might not be accurately described by \Cref{eq:2}, as also supported by our previous experimental results in \cite{martens_2021}. Furthermore, anti-angiogenic drugs are known to dramatically reduce vasogenic oedema without stopping tumour progression \cite{hawkins-daarud_2013}. Therefore, other MR sequences or modalities could be better suited for the initialisation and parameter estimation of reaction-diffusion glioma growth models. For instance, ADC maps derived from DW-MRI data could more accurately reflect tumour cell invasion as proposed in \cite{atuegwu_2012,hormuth_2021}. PET imaging with radio-labelled amino acids could also provide additional information to this extent as suggested in \cite{stockhammer_2008,lipkova_2019}.

Once the aforementioned prerequisites are met, our approach makes it possible to (i) extrapolate a whole brain tumour cell density distribution between and beyond the visible outlines of the tumour that is compatible with the reaction-diffusion model in \Cref{eq:4,eq:5,eq:6} and (ii) individually assess the values of the diffusivity and proliferation parameters of the model. Extrapolating tumour invasion is of utmost interest for radiotherapy planning since it would allow to define personalised margins which more accurately target the tumour while avoiding irradiation of the healthy tissues, as previously discussed in \cite{konukoglu_2010b, unkelbach_2014}. The independent assessment of the diffusivity and proliferation parameters of the model is for its part of great interest to better characterise the tumour \cite{le_2015}. The combination of both gives access to a fully personalised tool, initialised from clinical imaging data and allowing to anticipate the spatial-temporal growth of gliomas. Such a tool could for example be of considerable interest for dose fractionation optimisation in radiotherapy using a reinforcement learning approach as used in \cite{moreau_2021}. Furthermore, as it only depends on post-processed data (binary segmentations and a DTI-derived water diffusion tensor) and not on raw MR data, the proposed approach may be robustly extended to other scanners and centres. Besides, the method is also robust to variations in the imaging timing as variable imaging time intervals were considered for synthetic dataset generation, making it well adapted to the clinical reality.

The proposed method provided accurate estimations of the tumour cell distribution from only two imaging contours at a single time point, with a mean absolute error below \num[retain-unity-mantissa=false]{e-2} within the $c>0.01$ contour for the same tumours, as evaluated on 200 synthetic tumours grown over the brain domain of an unseen test subject. Our method also provided accurate estimates of the individual diffusivity and proliferation rate parameters of the model from three imaging contours extracted from two time points, with median relative errors of \SI{5.86}{\percent} and \SI{2.75}{\percent}, respectively (see \Cref{fig:7}), and strong concordance ($\mathrm{CCC} \geq 0.95)$ with the true parameter values (see \Cref{fig:8}). Furthermore, we showed that the spatio-temporal evolution of the tumour cell density distribution at later time points (\SI{90}{\day} and \SI{180}{\day} later) can be accurately captured from the estimated initial distribution and parameter values using the reaction-diffusion model. The ASSD between the true and estimated imaging contours obtained for threshold values of $c_1=0.80$ and $c_2=0.16$ were indeed found to be lower than or equal to the pixel spacing ($\SI{1}{\milli \metre} \times \SI{1}{\milli \metre} \times \SI{1}{\milli \metre}$) in most cases (see \Cref{fig:11}). Nevertheless, a loss of accuracy in the estimated tumour cell density over simulated time was observed (see \Cref{fig:10,fig:11,fig:9}), imputed to the accumulation of errors originating from uncertainties on the estimated model parameters values and initial condition. In particular, artefactual local maxima in the tumour cell densities distributions predicted by the CNN were found to give rise to new tumour foci over time. Post-processing steps were introduced to circumvent these effects (see \Cref{subsec:2_6}) but residual artefacts were still observed, resulting in a large Hausdorff distance though small ASSD values for a few isolated cases (see outliers in \Cref{fig:11}(a) and (b)). We also demonstrate the applicability of our proposed method to actual MR data of a GBM patient, for which we were able to reconstruct a tumour cell density distribution compatible with the imaging data. Nevertheless, the lack of biopsy samples combined with the multiple treatments undergone by the patient prevented the validation of the estimated distribution, which was left for a future prospective study.

As a future work, tumour-induced mass effect should be further integrated into the reaction-diffusion model since it is known to cause substantial deformations of the brain parenchyma as the tumour grows, which should also be taken into account for accurate treatment planning. Such effects have been previously considered \cite{clatz_2005, hogea_2007} but would introduce additional parameters to be assessed. Besides, transient brain deformations would hardly be integrated into a regular grid-based approach such as the finite difference method used in this work without loss of precision. A finite element formulation over an unstructured mesh could be used instead but would be much more computationally expensive---hence less suited for the generation of large high resolution datasets as the one described herein. Necrosis should also be taken into account by the model as proposed in \cite{swanson_2011, gu_2012}, which would have avoided the counter-intuitive correspondence between the hyper-dense ($c \sim 1$) region of the estimated tumour cell density distribution and the necrotic area visible on MRI in \Cref{fig:12}. Furthermore, the deep neural networks presented herein remain little flexible as they would need to be retrained if different imaging threshold values were considered, although transfer learning could be used to benefit from the lower-level features learned herein and avoid retraining the networks from scratch \cite{pan_2010}. Ultimately, the threshold values could be fed to the networks along with the binary contours, but would make the problem even more complex and would therefore require an even larger training dataset. Although real medical imaging data were used in this work, the validation of our approach still relied on healthy subject data. Therefore, the underlying hypothesis was made that the reaction-diffusion model defined by \Cref{eq:4,eq:5,eq:6} and used for tumour synthesis is indeed able to accurately capture the growth of real gliomas, which has never been extensively demonstrated so far to the best of our knowledge. Validation of our approach on actual glioma patient data should be further performed, but longitudinal imaging data with stereotactic biopsies of untreated glioma patients remain scarce. Including the effects of treatments into reaction-diffusion models have also been proposed previously \cite{woodward_1996, tracqui_1995, swanson_2002, hormuth_2021, rockne_2010} but again introduce additional parameters, increasing the complexity of the problem. Finally, the assumption was made throughout this work that the diffusivity and proliferation rate parameters of the model can be considered constant over time in untreated patients. The imaging time interval over which this assumption can reasonably hold should also be determined based on real glioma patient data. 

This work further demonstrates the ability of DCNNs to accurately approximate functions over complex domains and provides encouraging results towards the full personalisation of reaction-diffusion glioma growth models, which has remained an unsolved problem for decades.

\section{Conclusion}
We proposed a deep learning-based approach to address together the problems of estimating the initial condition and parameter values of a reaction-diffusion glioma growth model from patient magnetic resonance imaging data. We demonstrated the accuracy of our approach on synthetic tumours grown over actual brain domains of healthy volunteers. We also showed the applicability of our method on MR data of a real glioblastoma patient. Our promising results towards the full personalisation of glioma reaction-diffusion models may open up tremendous possibilities in the field.

\section*{Funding}
C.M. is funded by the Walloon Region, Belgium (PROTHER-WAL). C.D. is a senior research associate at F.R.S.-FNRS. The Department of Nuclear Medicine at H\^opital Erasme is supported by Association Vin\c{c}otte Nuclear (AVN), Fonds Erasme, and the Walloon Region (BioWin).

\section*{Acknowledgements}
The authors would like to thank the volunteers who kindly accepted to contribute to this study. The authors would also like to thank the technical, medical, and scientific staff in charge of image acquisition for the needs of this work.

\bibliographystyle{unsrt}
\bibliography{bibliography}

\appendix
\renewcommand{\theequation}{A\arabic{equation}}
\renewcommand{\thelstlisting}{A\arabic{lstlisting}}
\renewcommand{\thefigure}{A\arabic{figure}}
\renewcommand{\thetable}{A\arabic{table}}
\setcounter{equation}{0}
\setcounter{figure}{0}
\setcounter{lstnumber}{0}
\setcounter{table}{0}

\section{FSL DTI Data Processing}
\label{app:a}
The DICOM files of both acquisitions---with and without phase-encoding polarity inversion---were first converted into \lstinline{NIfTI} format using \lstinline{dcm2nii.exe} available in the MRIcroGL software (version 1.0.20180623) \cite{li_2016}. The 2 resulting series of 4 files with extensions \lstinline{.bvals}, \lstinline{.bvecs}, \lstinline{.json}, and \lstinline{.nii} were then renamed \lstinline{dti_pepolar_0} and \lstinline{dti_pepolar_1}, respectively for the acquisition with and without phase-encode polarity inversion. An acquisition parameter file \lstinline{acqparams.txt} specifying for each acquisition the phase-encode direction and the total readout time was then created and is provided in \Cref{lst:a_2}. For GE scanners, the total readout time $t_\text{read}$ [\si{\second}] is given by:
\begin{equation}
t_\text{read} = \left(n_y \, f_\text{acc} - 1\right) s_\text{echo} \times 10^{-6}
\label{eq:a_1}
\end{equation}
where $n_y$ is the acquisition matrix size along the phase-encode direction (DICOM tag (0018,1310)), $f_\text{acc}$ is the total phase acceleration factor given by the product of the ASSET and ARC factors (DICOM tag (0043,1083), ASSET/ARC), and $s_\text{echo}$ is the effective echo spacing (DICOM tag (0043,102C)) [\si{\micro \second}].

The Linux bash script used for DTI data processing using FSL (version 5.0.9-4) \cite{jenkinson_2012} is provided in \Cref{lst:a_1}. Lines 3--5 gather both $b=0$ volumes into a single 4D volume \lstinline{b0_merged.nii.gz}. Line 7 executes \lstinline{topup} on the merged volume with acquisition parameters in \lstinline{acqparams.txt}, which generates files \lstinline{b0_topup.nii.gz} (the corrected $b=0$ volumes), \lstinline{topup_results_fieldcoef.nii.gz} (the estimated susceptibility field), and \lstinline{topup_results_movpar.txt} (the estimated movement parameters). Line 9 computes the average of the two corrected volumes in \lstinline{b0_topup.nii.gz} and stores it as \lstinline{b0_topup_mean.nii.gz}. Line 10 computes a brain mask for \lstinline{b0_topup_mean.nii.gz} using \lstinline{bet}\cite{smith_2002} and generates files \lstinline{b0_topup_brain_mask.nii.gz} (the computed brain mask) and \lstinline{b0_topup_brain.nii.gz} (the masked volume). Lines 12--14 create a file specifying for each 3D volume in the 4D volume \lstinline{dti_pepolar_0.nii} (1 per diffusion direction), the corresponding acquisition parameters line in the \lstinline{acqparams.txt} file. Line 16 runs \lstinline{eddy} with the `replace outliers' option \lstinline{--repol} on \lstinline{dti_pepolar_0.nii} provided the previously detailed files and generates various result files including the corrected 4D volume \lstinline{dti_pepolar_0_eddy.nii.gz}. Line 8 finally runs \lstinline{dtifit} with the `save tensor' option \lstinline{--save_tensor} on the distortion-corrected diffusion data \lstinline{dti_pepolar_0_eddy.nii.gz}, which generates the fitted diffusion tensor file \lstinline{dti_pepolar_0_eddy_tensor.nii.gz} and various tensor-derived data.

\lstdefinestyle{bash}{
    backgroundcolor=\color{lightgray},
    commentstyle=\color{red}, 
    keywordstyle=\color{blue}, 
    language=bash, 
    morekeywords={fslroi, fslmerge, topup, fslmaths, bet, eddy_openmp, dtifit},
    numbers=left,
    classoffset = 1,
    otherkeywords = {topup=},
    morekeywords = {topup=},
    keywordstyle = {\color{black}},
    classoffset = 0
}

\begin{lstlisting}[backgroundcolor=\color{lightgray}, caption={Acquisition parameter file provided to \lstinline{topup}. The first three numbers of each line specify the phase-encode direction in the image coordinates system for the acquisition with (first line) and without (second line) phase-encore direction inversion. The last number of each line is the readout time $t_\text{read}$ given by \Cref{eq:a_1}.}, label={lst:a_2}, numbers=left]
0 1 0 0.04427
0 -1 0 0.04427
\end{lstlisting}

\begin{lstlisting}[caption={Linux bash script for DTI data processing using FSL.}, label={lst:a_1}, style=bash]
#!/bin/bash

fslroi dti_pepolar_1 b0_pepolar_1 0 1
fslroi dti_pepolar_0 b0_pepolar_0 0 1
fslmerge -t b0_merged b0_pepolar_1 b0_pepolar_0

topup --imain=b0_merged --datain=acqparams.txt --config=b02b0.cnf --out=topup_results --iout=b0_topup

fslmaths b0_topup -Tmean b0_topup_mean
bet b0_topup_mean b0_topup_brain -f 0.25 -m

idx=""
for ((i=1; i<=33; i+=1)); do idx="$idx 2"; done
echo $idx > index.txt

eddy_openmp --imain=dti_pepolar_0 --mask=b0_topup_brain_mask --acqp=acqparams.txt --index=index.txt --bvecs=dti_pepolar_0.bvecs --bvals=dti_pepolar_0.bvals --topup=topup_results --repol --out=dti_pepolar_0_eddy

dtifit --data=dti_pepolar_0_eddy --out=dti_pepolar_0_eddy --mask=b0_topup_brain_mask --bvecs=dti_pepolar_0_eddy.eddy_rotated_bvecs --bvals=dti_pepolar_0.bvals --wls --save_tensor
\end{lstlisting}

\section{Publicly Available Data}
\label{app:b}
The processed MR data of the 6 healthy volunteers used for this study are publicly available at \url{https://lisaserver.ulb.ac.be/owncloud/index.php/s/KwEPG65gh1U7xNM}. For each volunteer, a directory is provided containing the \lstinline{brain_domain.mha}, \lstinline{unit_diffusion_tensor.mha}, and \lstinline{unit_proliferation_rate.mha} files in \lstinline{MetaImage} format. \lstinline{brain_domain.mha} contains the segmented brain map as derived in \Cref{subsubsec:2_3_4}, stored as an \lstinline{unsigned short} 3D image. Values 0, 2, 3, and 4 respectively correspond to background, cerebrospinal fluid, grey matter, and white matter voxels. \lstinline{unit_diffusion_tensor.mha} contains the 6 independent components $d_{xx}$, $d_{xy}$, $d_{xz}$, $d_{yy}$, $d_{yz}$, and $d_{zz}$ of the unit tumour cell diffusion tensor as derived in \Cref{subsubsec:2_3_6}, stored as a \lstinline{double} 4D image. \lstinline{unit_proliferation_rate.mha} contains a dummy proliferation rate field stored as a \lstinline{double} 3D image with value 1 for white and grey matter voxels and 0 elsewhere, to be scaled by the proliferation rate $\rho$ and fed to the \lstinline{tgstkFiniteDifferenceReactionDiffusionTumourGrowthFilter} (see \url{https://cormarte.github.io/tgstk/html/classtgstk_finite_difference_reaction_diffusion_tumour_growth_filter.html}).

\section{Additional Results}
\label{app:c}
\begin{figure}[ht!]
    \centering
    \includegraphics[width=0.75\textwidth]{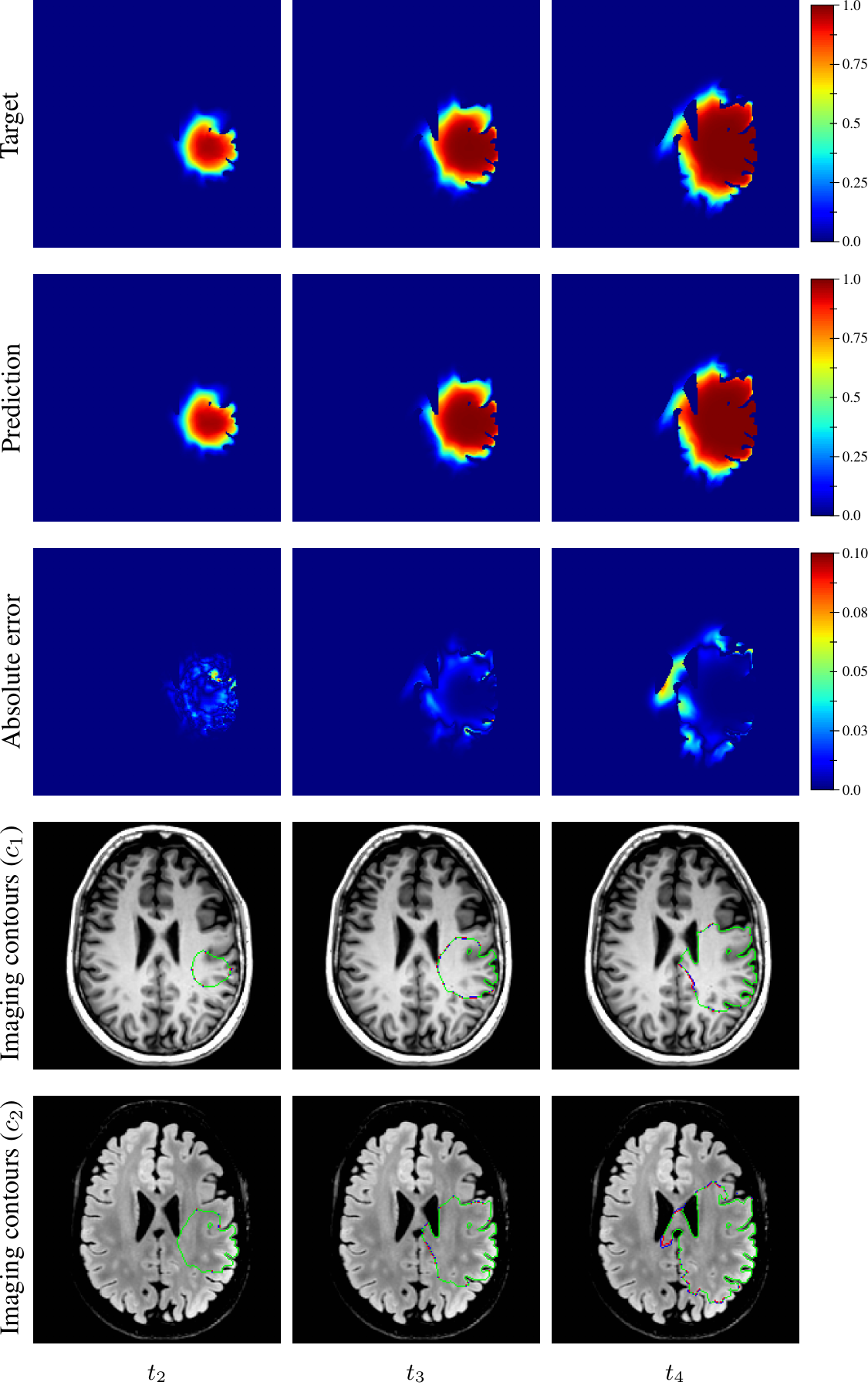}
    \caption{Example of true (1\textsuperscript{st} row) and estimated (2\textsuperscript{nd} row) tumour cell density distributions at times $t_{2-4}$ (1\textsuperscript{st} to 3\textsuperscript{rd} column, axial slices) along with the corresponding absolute error maps (3\textsuperscript{rd} row) for a test tumour ($d=\SI{46.20}{\square \milli \meter \per \year}$, $\rho=\SI{12.92}{\per \year}$, $t_1=\SI{130}{\day}$, $t_2=\SI{268}{\day}$). The imaging contours for threshold values $c_1=0.80$ and $c_2=0.16$ superimposed to the T1 and T2 FLAIR image are depicted in the 4\textsuperscript{th} and 5\textsuperscript{th} rows, respectively. The blue, red, and green segments respectively correspond to the target, prediction, and overlapping contour voxels.}
    \label{fig:a1}
\end{figure}

\begin{figure}[ht!]
    \centering
    \includegraphics[width=0.75\textwidth]{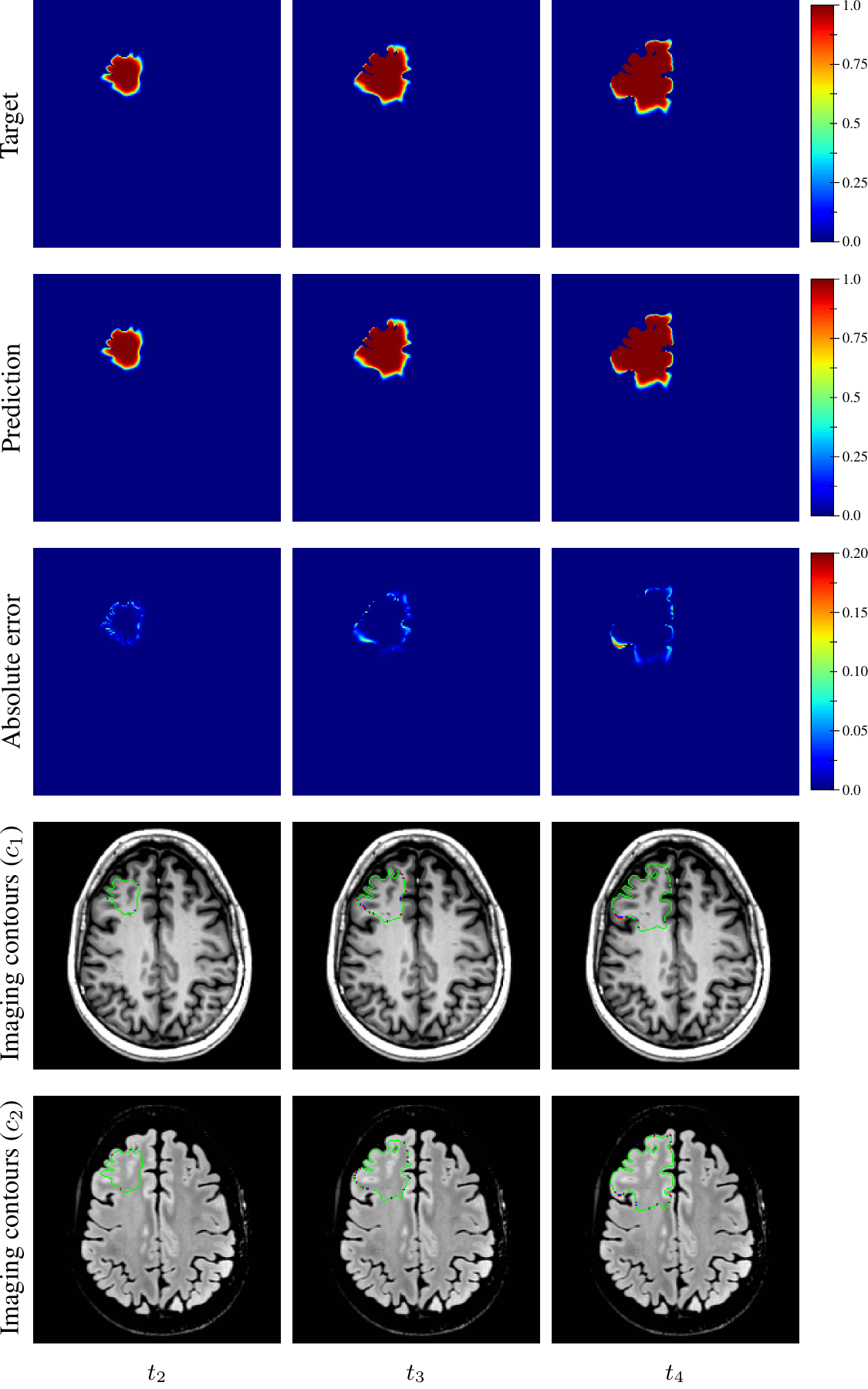}
    \caption{Example of true (1\textsuperscript{st} row) and estimated (2\textsuperscript{nd} row) tumour cell density distributions at times $t_{2-4}$ (1\textsuperscript{st} to 3\textsuperscript{rd} column, axial slices) along with the corresponding absolute error maps (3\textsuperscript{rd} row) for a test tumour ($d=\SI{9.52}{\square \milli \meter \per \year}$, $\rho=\SI{25.77}{\per \year}$, $t_1=\SI{94}{\day}$, $t_2=\SI{225}{\day}$). The imaging contours for threshold values $c_1=0.80$ and $c_2=0.16$ superimposed to the T1 and T2 FLAIR image are depicted in the 4\textsuperscript{th} and 5\textsuperscript{th} rows, respectively. The blue, red, and green segments respectively correspond to the target, prediction, and overlapping contour voxels.}
    \label{fig:a2}
\end{figure}

Additional examples of true and estimated tumour cell density distributions from the test set at times $t_{2-4}$ obtained as described in \Cref{subsec:2_6} are depicted in \Cref{fig:a1,fig:a2}, along with the corresponding absolute error maps and imaging contours.

\end{document}